\documentclass[conference]{IEEEtran}
\IEEEoverridecommandlockouts
\usepackage{cite}
\usepackage{amsmath,amssymb,amsfonts}
\usepackage{algorithm}
\usepackage{threeparttable}
\usepackage{dsfont}
\usepackage{algpseudocode}   
\usepackage{graphicx}
\usepackage{tabularx}
\usepackage{textcomp}
\usepackage{booktabs}
\usepackage{pifont}
\usepackage{xcolor}
\usepackage{multirow}
\usepackage{makecell}
\usepackage{array}
\def\BibTeX{{\rm B\kern-.05em{\sc i\kern-.025em b}\kern-.08em
    T\kern-.1667em\lower.7ex\hbox{E}\kern-.125emX}}
\begin{document}

\title{An Empirical Study on the Impact of Normalized Use‑Case Specifications on Traceability}
\author{
    \IEEEauthorblockN{
        Luoyuan Shi\IEEEauthorrefmark{1}\IEEEauthorrefmark{2},
        Yuanzhao Zhai\IEEEauthorrefmark{1}\IEEEauthorrefmark{2},
        Dawei Feng\IEEEauthorrefmark{1}\IEEEauthorrefmark{2},
        Jialin Zhao\IEEEauthorrefmark{1}\IEEEauthorrefmark{2},
        Zhaoxie Xu\IEEEauthorrefmark{1}\IEEEauthorrefmark{2},
        Bo Ding\IEEEauthorrefmark{1}\IEEEauthorrefmark{2},
        Huaimin Wang\IEEEauthorrefmark{1}\IEEEauthorrefmark{2}
    }
    \IEEEauthorblockA{\IEEEauthorrefmark{1}College of Computer Science and Technology, National University of Defense Technology, Changsha, China}
    \IEEEauthorblockA{\IEEEauthorrefmark{2}State Key Laboratory of Complex \& Critical Software Environment, Changsha, China}
    \IEEEauthorblockA{shiluoyuan1995@126.com}
}
\maketitle
\begin{abstract}
Traceability link recovery between requirements and code is critical for software quality assurance and evolution analysis. Although traceability technologies have made considerable progress, there exists a huge semantic gap between ambiguous natural‑language requirements and precise source code, which constitutes the fundamental obstacle to recovering traceability links between requirements and code. Most existing approaches focus on improving traceability techniques for requirements and code while ignoring the quality of requirement descriptions themselves. Such technology‑oriented optimization that overlooks source‑data quality makes it difficult to fundamentally bridge the semantic gap. To address this issue, this study proposes to start from requirements: adopting controlled‑natural‑language specifications to unify text structures and filter noise, and leveraging large language models to complement implicit domain context. This enhances the discriminability of semantic representations of requirements in vector space and narrows the semantic gap. We first select four public datasets and employ large language models together with prompt engineering to decompose requirements and transform them into normalized use‑case specifications. We then systematically evaluate the impact of normalized use‑case specifications on the performance of requirement‑to‑code traceability link recovery using two representative frameworks. The empirical results demonstrate that normalized use‑case specifications can effectively reduce the semantic gap for original requirements with semantic ambiguity and yield consistent performance gains across different traceability frameworks. Meanwhile, this study identifies the applicability boundary of the proposed method: for high‑quality requirements whose vocabulary already highly aligns with underlying source code, over‑normalization may dilute feature words and degrade performance. Therefore, requirement quality assessment is recommended prior to practical deployment. This research verifies that requirement normalization at the source is an effective strategy to bridge the semantic gap and boost traceability performance, offering new perspectives for traceability link‑recovery research.
\end{abstract}


\section{Introduction}
Traceability between software requirements and code is a core capability for guaranteeing software quality, supporting change‑impact analysis and compliance verification\cite{tao2022survey}. In modern complex software systems, effective requirement traceability not only ensures consistency between system implementations and user intentions but also significantly reduces maintenance costs. Studies show that each additional untraced requirement item raises the average software defect rate by approximately 2.3\%. Nevertheless, with the explosive growth of software scale, manually maintaining traceability links can no longer satisfy high‑frequency, low‑latency delivery demands under agile iteration. Accordingly, efficient Traceability Link Recovery (TLR) techniques have become a shared goal for both academia and industry.

Currently, TLR tasks face a core bottleneck: the semantic gap between ambiguous requirements and precise code. Specifically, original requirements focus on high‑level descriptions of business logic, whereas source code targets concrete functional implementations, bringing substantial asymmetry in abstraction levels and vocabulary usage. For instance, a simple requirement such as ``payment function'' may correspond to multiple complex code logics involving encryption, interfaces and databases. Such one‑to‑many mappings greatly increase the difficulty of link recovery.

Existing research mostly concentrates on traceability‑technique optimization, striving to improve requirement‑to‑code traceability‑recovery accuracy via sophisticated deep‑learning models or Retrieval‑Augmented Generation (RAG). However, such technology‑centric optimization that disregards source‑data quality cannot fundamentally bridge the semantic gap. Taking the iTrust dataset as an example, when tracing requirement UC10E1 using information‑retrieval‑based traceability, \textit{AddPatientValidator.java} contains abundant frequent domain terms such as patient, validate, and invalid that resemble UC10E1. Despite semantic misalignment, lexical co‑occurrence leads to false‑positive outputs. In contrast, \textit{GetUserNameAction.java} is semantically compatible yet lacks surface‑level lexical overlap with the requirement (e.g., identifier inputMID), and is consequently missed by the system without auxiliary information.

Motivated by the above observations, this study aims to narrow the semantic gap starting from the requirement side. Controlled Natural Language (CNL) delivers prominent advantages for processing unstandardized texts owing to its predefined structured templates and constrained vocabulary. By mapping unstructured natural‑language texts into uniformly formatted, precise descriptions with rich semantic information, CNL eliminates lexical ambiguity and improves requirement consistency. It provides high‑signal‑to‑noise‑ratio inputs for Traceability Link Recovery (TLR), and its normalized features also improve automation efficiency as well as the accuracy of context‑semantic understanding by Large Language Models (LLMs). In this work, four public datasets are adopted. LLMs are utilized to reconstruct original requirements into use‑case specifications conforming to verbose‑style Controlled Natural Language (CNL‑B). Afterwards, we evaluate how normalized use‑cases affect traceability performance under two frameworks: the information‑retrieval‑based Fine‑grained Traceability Link Recovery (FTLR)\cite{hey2021improving}, and the RAG‑based Linking Software System Artifacts (LISSA)\cite{fuchss2025lissa}.

Our main contributions are summarized as follows:
\begin{enumerate}
    \item Benchmark dataset construction: We build and release a high‑quality triple dataset consisting of original requirements, normalized use‑cases and source code to facilitate future research.
    \item Automated use‑case‑specification generation framework: We propose an LLM‑driven automated generation approach for CNL‑B‑compliant normalized use‑case specifications, covering requirement decomposition, structured generation and quality validation.
    \item Empirical findings: We quantitatively verify the benefits of normalized use‑case specifications in narrowing semantic gaps and boosting TLR performance (especially F1‑score).
    \item Best practices for specification generation: To address coarse‑grained and mixed‑function original requirements that hinder precise traceability, we introduce a requirement‑decomposition mechanism. Two decomposition strategies (business‑logic‑oriented versus technical‑layer‑oriented) are compared regarding specification generation and traceability performance. The results indicate that business‑logic‑oriented decomposition preserves complete business logic while producing fewer normalized use‑case specifications. It reduces computational overhead and benefits both retrieval matching and human comprehension.
\end{enumerate}

The remainder of this paper is organized as follows. Section 2 reviews related work on requirement‑to‑code traceability link recovery. Section 3 elaborates our research methodology, including dataset collection, processing and quality assurance. Section 4 validates the effectiveness of normalized use‑case specifications via comparative experiments. Section 5 analyzes factors threatening validity. Section 6 provides discussions. Section 7 concludes this paper.

\section{Related Work}
The domains relevant to this study mainly include requirement‑to‑code traceability link recovery and standardized representation approaches for use‑case specifications.

\subsection{Traceability Link Recovery (TLR)}
TLR aims to identify correspondences between requirements and software artifacts such as source code. Effective requirement traceability is essential for software quality. As mentioned earlier, missing traceability links directly lead to a notable rise in software defect rates\cite{rempel2016preventing}. Traceability link recovery techniques can be broadly divided into traditional methods and learning‑based methods.

Traditional methods are primarily built upon information retrieval techniques. Candidate links are identified by computing textual similarity across different software artifacts, under the core assumption that semantic similarity among artifacts positively correlates with traceability\cite{tao2022survey}. Early studies widely adopted techniques including the Vector Space Model (VSM) and Latent Semantic Indexing (LSI)\cite{ballarin2023influence} to detect candidate links. To mitigate the prevalent lexical mismatch between requirements and code, researchers have extended basic IR models. For example, Díaz et al. proposed TYRION, which alleviates lexical mismatch by combining code author information with information retrieval\cite{diaz2013using}. Hayes et al. improved the vector‑space model with external synonym lexicons to boost recall and precision on the NASA MODIS dataset\cite{hayes2003improving}. De Lucia et al. readjusted term weights by introducing smoothing filters to remove “common noise” within homogeneous software artifacts such as code classes\cite{de2011improving}. Building upon this, they further presented a hybrid framework that combines orthogonal IR models (e.g., RTM and VSM/JS) via affine transformation, leveraging complementary properties across models to enhance traceability accuracy\cite{gethers2011integrating}. Moreover, fine‑graining traceability elements constitutes another important direction for improving TLR performance. Methods such as FTLR adopt fastText word embeddings and WMD to compute similarity between sentence‑level requirements and code methods, yielding better experimental results than conventional coarse‑grained document‑level matching\cite{hey2021improving}.

Learning‑based methods enhance traceability accuracy by capturing feature patterns. Early work leveraged classic machine‑learning algorithms such as SVM and decision trees, as well as deep‑learning architectures including CNN and RNN, to learn nonlinear mappings among artifacts\cite{guo2017semantically}. While these approaches achieve promising results on specific datasets, their heavy reliance on high‑quality labeled data limits their generalization capability in agile projects with scarce annotations. In recent years, LLM‑driven TLR has attracted substantial research attention. In terms of model fine‑tuning, specialized models such as NoRBERT are adopted to detect and filter non‑functional noise within requirements, supplying higher‑quality inputs for traceability algorithms\cite{hey2024requirements}. Chuyan Ge et al. optimized LLaMA variants of different sizes through data augmentation, prompting and fine‑tuning. Their approach outperforms traditional IR and deep‑learning methods on cross‑level requirement traceability tasks, yet exposes limitations such as sensitivity to data augmentation and dependencies on long‑text processing\cite{ge2025cross}. To better unlock the reasoning potential of LLMs, researchers have intensively investigated prompt engineering for traceability tasks. For instance, Alberto D. Rodríguez et al. revealed how minor prompt modifications influence traceability outputs, and proposed prompting strategies with chain‑of‑reasoning and multi‑view classification to improve model performance across diverse datasets\cite{rodriguez2023prompts}. On this basis, the LISSA framework further integrates Retrieval‑Augmented Generation (RAG). It performs retrieval over candidate links followed by secondary reasoning judgment to further boost traceability performance\cite{fuchss2025lissa}.

Nevertheless, existing research predominantly focuses on optimizing downstream traceability algorithms while overlooking the root‑cause problem of poor original‑requirement quality. Specifically, traditional IR methods heavily depend on lexical overlap and struggle to bridge the semantic gap between requirements and code under common agile‑development challenges such as terminology mismatch and text sparsity. Learning‑based approaches (including LLM‑based solutions) intrinsically require high‑quality labeled data. When input requirement descriptions are ambiguous and unstandardized, their performance is highly vulnerable to noise and suffers degraded generalization. Accordingly, the effectiveness of both lexicon‑dependent IR methods and annotation‑reliant learning methods is constrained by input‑data ambiguity and inconsistency. Motivated by this observation, this study targets the requirement side. By introducing normalized use‑case specifications, we improve the semantic clarity and consistency of requirements and provide higher‑quality inputs for subsequent traceability tasks.

\subsection{Controlled Natural Language}
To reduce the semantic gap between unstructured requirements and code implementations, structured requirement representations such as use‑case specifications have long been regarded as potential “semantic bridges”. Use‑case specifications can deliver richer semantic information than raw requirements, including concrete operational workflows and constraints. Transforming raw requirements into use‑case specifications improves requirement precision, analyzability and traceability while preserving readability\cite{hey2024requirements,darif2026controlled}. Studies have shown that writing use‑case specifications under CNL standards effectively reduces ambiguity, enhances requirement consistency, and lays a foundation for automated verification and test generation\cite{da2021linguistic}.

Nevertheless, existing CNL‑related research and applications still face two major challenges. First, from the research‑perspective, most existing work focuses on improving intrinsic requirement quality or downstream tasks such as test generation\cite{wang2020automatic}. Systematical empirical investigations into its effectiveness for requirement‑to‑code traceability remain scarce. Second, there are constraints imposed by production costs. Manually authoring high‑quality use‑case specifications incurs substantial overhead, which fundamentally conflicts with the rapid‑iteration paradigm advocated by prevalent agile development\cite{maro2022tracimo}. Existing use‑case generation tools (e.g., the RSL editor) still heavily rely on manual analysis and human annotation\cite{da2021linguistic}. The lack of automatic generation approaches limits their adoption in large‑scale projects.

Targeting the above challenges, this study adopts the verbose‑style Controlled Natural Language (CNL‑B) proposed by Alberto Silva\cite{da2021linguistic} as the specification standard for use‑case specifications. With predefined structured templates and constrained vocabularies, CNL‑B converts loose, ambiguous natural‑language requirements into standardized descriptions with uniform formatting, precise wording and complete information\cite{da2021linguistic}. Three major considerations motivate our selection of CNL‑B. First, logical compatibility: CNL‑B emphasizes the “subject‑action‑object” structure, which shares natural structural similarity with the “object‑method‑parameter” logic in source code. Second, purification and standardization: compared with unstructured requirements, CNL‑B filters descriptive noise to a certain extent and guarantees that each use‑case step constitutes a distinct functional unit. Third, automation‑friendliness: the well‑defined syntax of CNL‑B can be readily understood by LLMs, enabling automatic transformation from raw requirements to normalized use‑cases and substantially reducing human labor.

\section{Research Methodology}
\subsection{Research Questions}
To evaluate how CNL‑B‑compliant use‑case‑specification datasets affect requirement‑to‑code traceability performance under the IR‑based FTLR framework and the RAG‑based LISSA framework, we formulate the following research questions:

\noindent \textbf{RQ1}: To what extent do automatically generated normalized use‑case specifications preserve consistency with the original requirements?

This research question validates whether LLMs can accurately and completely inherit the semantic logic of original requirements while complying with CNL‑B formatting constraints. At the generation stage, prompt optimization and cross‑model validation are adopted to ensure generation stability. At the evaluation stage, a manual auditing scheme is applied to quantitatively score generated specifications across three dimensions: accuracy, completeness, and consistency. Moreover, Cohen's Kappa coefficient is introduced to assess inter‑annotator agreement, providing a reliable scientific benchmark for the quality of the generated dataset.

\noindent \textbf{RQ2}: Compared with original requirements, to what extent can normalized use‑case specifications improve traceability performance (F1‑score) under different traceability frameworks (i.e., IR‑based framework and RAG‑LLM framework)?

Under two representative approaches, namely FTLR and LISSA, this research question compares the performance of original requirements and CNL‑B‑compliant use‑case specifications under varying granularities and different similarity computation methods. Evaluated metrics include F1‑score, Precision, and Recall.

\noindent \textbf{RQ3}: What are the differences in the impacts exerted by different requirement‑decomposition strategies on use‑case‑specification generation and traceability performance?

By comparing two requirement‑decomposition strategies (business‑logic‑oriented versus technical‑layer‑oriented), this research question investigates their influences on use‑case‑specification generation and how such influences propagate to traceability tasks, so as to identify which decomposition strategy is more beneficial for traceability link recovery.

\subsection{Dataset Collection}
Four open‑source requirement‑to‑code traceability datasets are selected in this study: eANCI, eTour, iTrust and LibEST. These datasets cover diverse domains including government affairs, tourism, healthcare, and cybersecurity, adopt two mainstream programming languages (Java and C), and contain requirement artifacts in template‑based use‑cases, free‑form natural text, and event flows. To guarantee experimental rigor while reducing experimental overhead, we only generate CNL‑B‑compliant use‑case specifications via the LLM‑based automatic framework for requirements with ground‑truth traceability links in the four datasets. Due to varying granularity of original requirements, one original requirement may yield 1 to 6 relatively independent normalized use‑case specifications. Detailed statistics are listed in Table \ref{tab:dataset}.

\begin{table*}[htbp]
\centering
\caption{Experimental Datasets}
\resizebox{\linewidth}{!}{
\begin{tabular}{lcccccccccc}
\toprule
Dataset & Programming Language & Dataset Source & Requirement Type & Language & Domain & Requirement Files & Traceability Links & Requirements with Traceability Links & CNL‑B Use‑Case Specifications \\
\midrule
eANCI & JAVA & CoEST & Use‑case Template & Italian (automatically translated into English via Google Translate API) & Government System & 139 & 567 & 39 & 43 \\
eTour & JAVA & CoEST & Use‑case Template & English & Tourism System & 58 & 309 & 57 & 66 \\
iTrust & JAVA & CoEST & Event‑flow Only & English & Healthcare & 131 & 286 &105& 175 \\
LibEST & C & Moran et al.\cite{moran2020improving} & Natural Text & English & Cybersecurity & 52 & 204 & 47 & 71 \\
\bottomrule
\end{tabular}
}
\vspace{0.5em}
\footnotesize CoEST: Center of Excellence for Software \& Systems Traceability
\label{tab:dataset}
\end{table*}

\subsection{Evaluation Metrics}
This paper adopts widely‑used evaluation metrics in the requirement traceability community\cite{koboyatshwene2025requirements}, namely Recall, Precision, and F1‑score. Recall measures the completeness of traceability links identified by automatic algorithms, i.e., the percentage of correctly detected traceability links among all ground‑truth traceability links. The F1‑score is the harmonic mean of Precision and Recall. The three metrics are computed from Equation (1) to Equation (3):
\begin{align}
\mathrm{Recall} &= \frac{\mathrm{TP}}{\mathrm{TP}+\mathrm{FN}} \\
\mathrm{Precision} &= \frac{\mathrm{TP}}{\mathrm{TP}+\mathrm{FP}} \\
\mathrm{F1\text{-score}} &= \frac{2\cdot \mathrm{Precision} \cdot \mathrm{Recall}}{\mathrm{Precision} + \mathrm{Recall}}
\end{align}

Where:
True Positive (TP): number of samples correctly identified as possessing traceability links;
True Negative (TN): number of samples correctly identified as having no traceability links;
False Positive (FP): number of negative samples misclassified as having traceability links;
False Negative (FN): number of positive samples misclassified as having no traceability links.

\subsection{LLM‑based Automatic Generation of CNL‑B‑Compliant Normalized Use‑Case Specifications}
\begin{figure*}[t]
\centering
\includegraphics[width=0.9\linewidth]{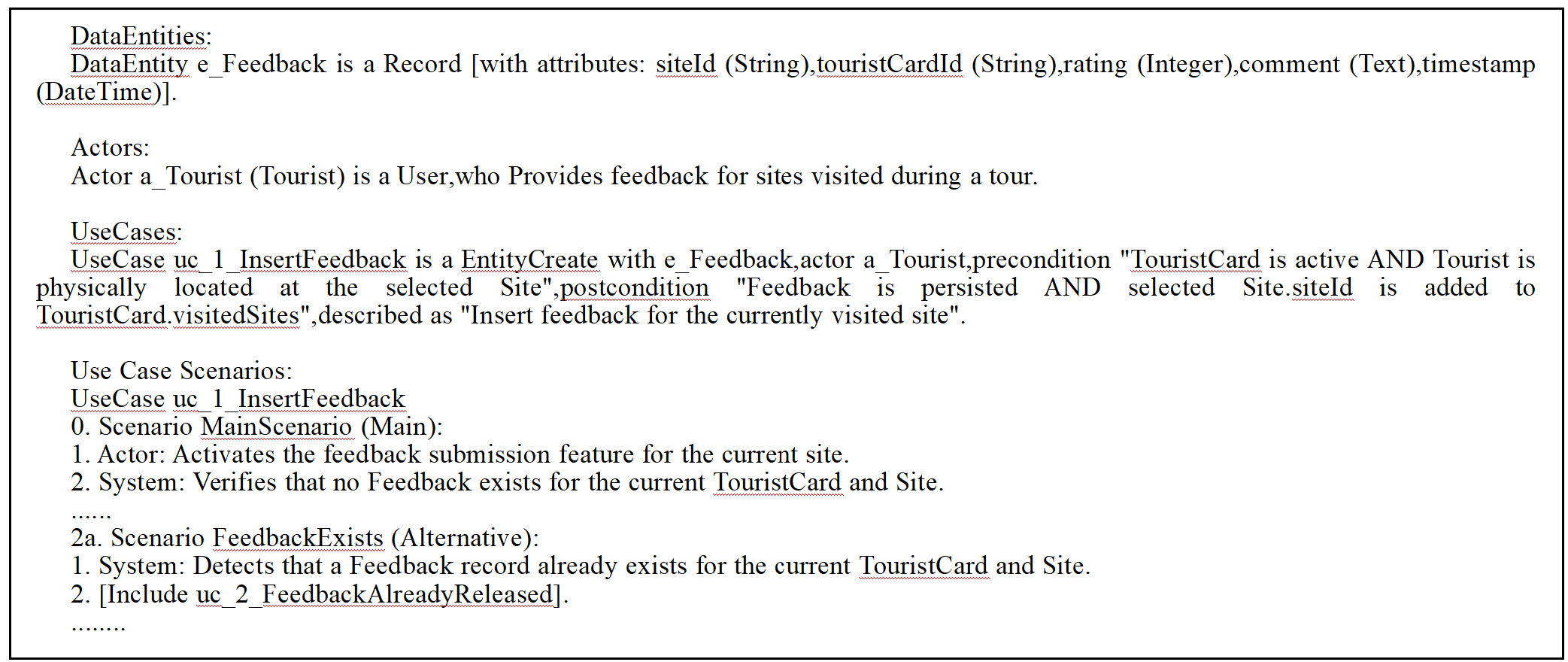}
\caption{Example of CNL‑B‑Compliant Use‑Case Specification}
\label{fig:1}
\end{figure*}

To quantitatively evaluate the potential of CNL‑B normalized use‑case specifications (see Figure \ref{fig:1}) for requirement‑to‑code TLR, two generation strategies are designed in this study:
\begin{enumerate}
    \item Requirement‑driven approach (detailed in this section): generates normalized use‑case specifications solely from original requirements. It simulates real‑world industrial scenarios where prior knowledge such as target source code and ground‑truth links is unavailable. This serves as our primary generation strategy.
    \item Code‑augmented approach (Appendix \ref{app:code_enhance}): leverages ground‑truth linked code artifacts to assist generation and serves as a comparison baseline. This variant explores whether injecting low‑level implementation details into use‑case specifications with access to linked‑code context can further bridge the semantic gap, or instead trigger feature dilution caused by abstraction‑level mismatch.
\end{enumerate}

\begin{figure*}[t]
\centering
\includegraphics[width=0.9\linewidth]{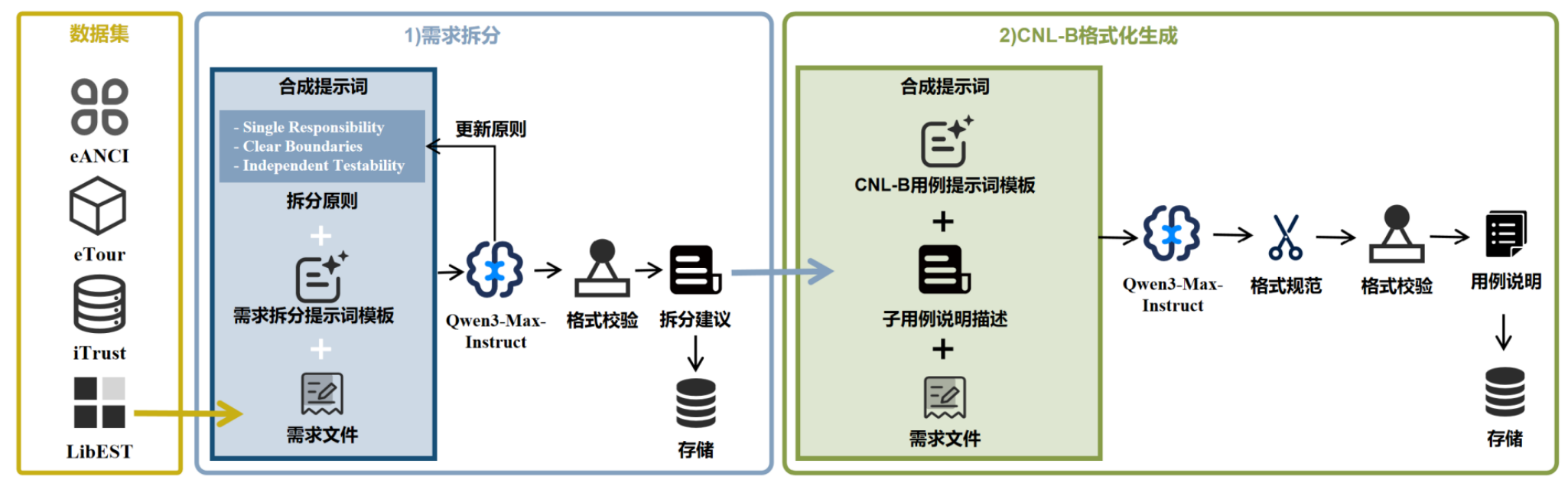}
\caption{LLM‑based Automatic Generation of CNL‑B‑Compliant Normalized Use‑Case Specifications}
\label{fig:2}
\end{figure*}

This section first introduces the requirement‑driven approach: the LLM‑based automatic generation method for CNL‑B‑compliant normalized use‑case specifications (see Figure \ref{fig:2}). This approach alleviates manual dependency in conventional use‑case authoring and mitigates subjective biases over use‑case granularity and format consistency. The framework consists of two core phases: 1) requirement decomposition: split original requirements into independent sub‑requirements; 2) CNL‑B formatting generation: transform sub‑requirements into standardized CNL‑B representations.

The use‑case generation task is built upon the Qwen3‑Max‑Instruct model, whose performance heavily relies on prompt strategy design. Accordingly, Chain‑of‑Thought (CoT) prompting is adopted to enhance reasoning stability and output standardization.

\subsubsection{Requirement Decomposition Algorithm}
The core task of this phase is to produce sub‑requirement decomposition suggestions $\mathcal{A}_{analysis\_results}$ for original requirements by exploiting LLM reasoning under predefined decomposition guidelines. We first conduct evaluations over decomposition strategies, and construct prompts based on the optimal strategy to perform requirement decomposition. Furthermore, a requirement decomposition algorithm is designed and implemented, as shown in Algorithm \ref{alg:decomposition_v1}. The algorithm adopts an iterative loop mechanism. In each iteration, the system assembles decomposition principles, prompt templates and input requirement files, feeds them into Qwen3‑Max‑Instruct to generate decomposition suggestions. Outputs are stored after passing format validation.

\begin{algorithm}[!htbp]
\centering
\begin{algorithmic}[1]
\Require Original requirement dataset $\mathcal{S}_{data}$, Large Language Model $\text{Qwen3‑Max‑Instruct}$, Initial decomposition principles $\mathcal{P}$
\Ensure Requirement decomposition analysis results $\mathcal{A}_{\text{analysis\_results}}$

\For{\textbf{each} requirement file $req_i \in \mathcal{S}_{data}$}
    \State $Prompt \gets \text{AssemblePrompt}(req_i, \mathcal{P})$ \Comment{Assemble principles, template and input file}
    \State $Response \gets \text{Qwen3‑Max‑Instruct}.\text{Generate}(Prompt)$
    
    \If{$\text{CheckFormat}(Response)$} \Comment{Format validation}
        \State $Result_i, \mathcal{P}_{\text{new}} \gets \text{Parse}(Response)$
        \State $\mathcal{P} \gets \text{Update}(\mathcal{P}, \mathcal{P}_{\text{new}})$ \Comment{Principle evolution: absorb decomposition features and update in real‑time}
        \State $\mathcal{A}_{\text{analysis\_results}}[req_i] \gets Result_i$
    \EndIf
\EndFor

\Return $\mathcal{A}_{\text{analysis\_results}}$
\end{algorithmic}
\caption{Requirement Decomposition Algorithm}
\label{alg:decomposition_v1}
\end{algorithm}

This algorithm introduces a principle‑evolution mechanism. Each iteration absorbs decomposition characteristics summarized from the current task and updates rules dynamically, improving model adaptability to diverse requirement texts.

Three core decomposition principles are initialized:
\begin{itemize}
    \item Single responsibility: clear business operations initiated by actors;
    \item Clear boundaries: well‑defined inputs, processing workflows, and outputs;
    \item Independently testable: testable without relying on other sub‑requirements.
\end{itemize}

\subsubsection{CNL‑B Use‑Case Specification Generation Algorithm}

This phase generates CNL‑B‑compliant use‑case specifications from the sub‑requirement decomposition suggestions $\mathcal{A}_{\text{analysis\_results}}$ produced by Algorithm \ref{alg:decomposition_v1}. The generation procedure is illustrated in Algorithm \ref{alg:cnl_generation_v1}.

\begin{algorithm}[!htbp]
\centering
\begin{algorithmic}[1]
\Require Sub‑requirement decomposition suggestions $\mathcal{A}_{\text{analysis\_results}}$, original requirement dataset $\mathcal{S}_{data}$, Large Language Model $\text{Qwen3‑Max‑Instruct}$, CNL‑B prompt template $\text{Template}_{\text{CNL‑B}}$
\Ensure Result set of CNL‑B use‑case specifications $\mathcal{R}_{\text{cnl}}$

\State $\mathcal{R}_{\text{cnl}} \gets \emptyset$

\For{\textbf{each} sub‑requirement $sub\_req_i \in \mathcal{A}_{\text{analysis\_results}}$}
    \State $is\_passed \gets \mathbf{False}$
    \While{\textbf{not} $is\_passed$}
        \State $Prompt \gets \text{ConstructPrompt}(sub\_req_i, \mathcal{S}_{data},$
        \State \quad $\text{Template}_{\text{CNL‑B}})$
        \State $Response \gets \text{Qwen3‑Max‑Instruct}.\text{Generate}($
        \State \quad $Prompt)$

        \State $\text{Standardized\_Doc} \gets \text{FormatNormalize}(Response)$ \Comment{Format normalization processing}

        \If{$\text{CheckModules}(\text{Standardized\_Doc})$} \Comment{Validate modules such as Actors, UseCases, Scenarios}
            \State $is\_passed \gets \mathbf{True}$
            \State $\mathcal{R}_{\text{cnl}} \gets \mathcal{R}_{\text{cnl}} \cup \{\text{Standardized\_Doc}\}$
        \Else
            \State \textbf{continue} \Comment{Validation failed, trigger re‑generation}
        \EndIf
    \EndWhile
\EndFor

\Return $\mathcal{R}_{\text{cnl}}$
\end{algorithmic}
\caption{CNL‑B Use‑Case Specification Generation Algorithm}
\label{alg:cnl_generation_v1}
\end{algorithm}

First, the system assembles structured prompts from sub‑requirement descriptions, original requirements, and the CNL‑B use‑case prompt template according to decomposition suggestions. The Qwen3‑Max‑Instruct model is invoked to produce CNL‑B‑compliant use‑case specifications. To guarantee syntactic and structural compliance of the specifications, the system sequentially performs format normalization and format checking. Core components including Actors, Use Cases, and Use Case Scenarios are primarily validated. Documents passing validation are stored; otherwise, a re‑generation instruction is triggered.

\subsection{Experimental Setup}
To systematically evaluate the effectiveness and generality of normalized use‑case specifications under different technical architectures, this study selects FTLR\cite{hey2021improving} and LISSA\cite{fuchss2025lissa} as baseline experimental frameworks. FTLR represents the traditional information‑retrieval‑based traceability paradigm, whereas LISSA stands for the novel traceability paradigm built upon retrieval‑augmented generation and large language models. Both are open‑source projects and have been repeatedly cited in related research, ensuring experimental reproducibility and comparability.

\begin{figure}[t]
\centering
\includegraphics[width=0.9\linewidth]{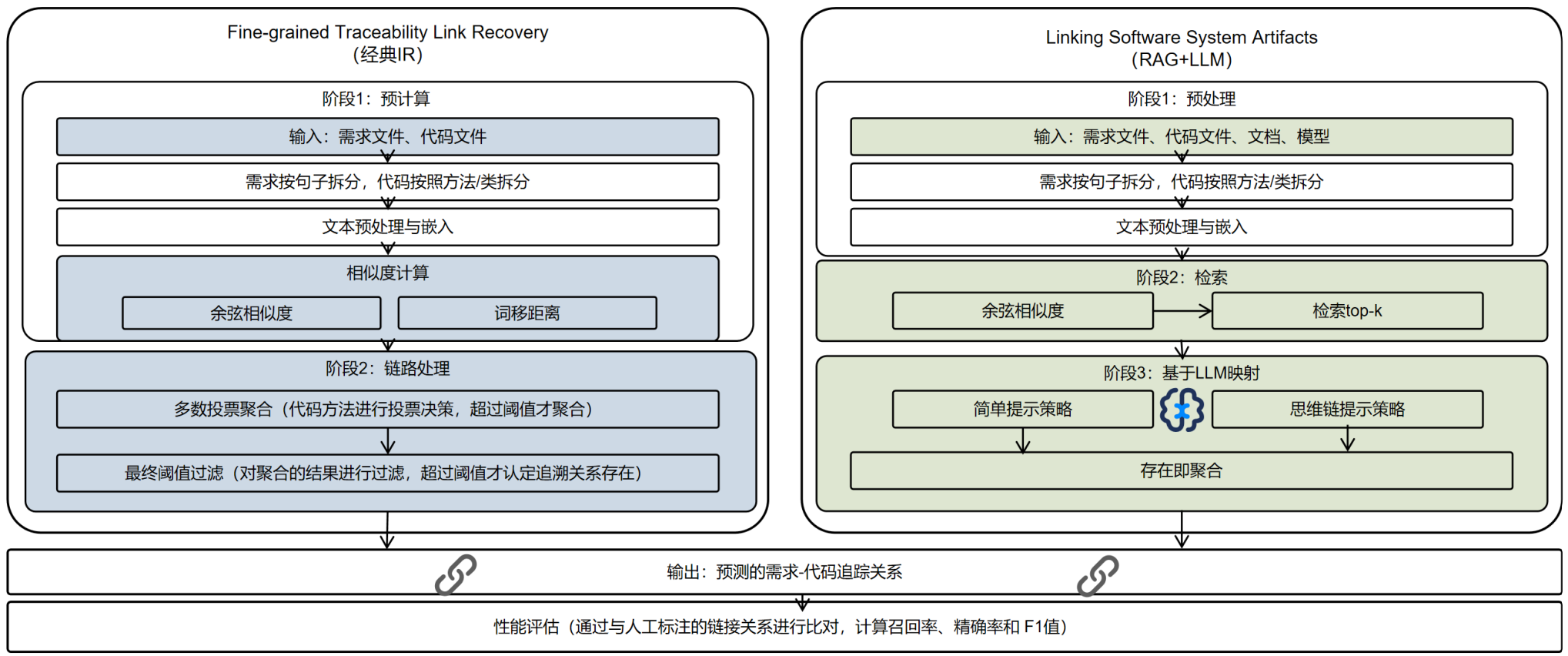}
\caption{Comparison between FTLR and LISSA Approaches}
\label{fig:fig5}
\end{figure}

Figure \ref{fig:fig5} illustrates the comparison between the FTLR and LISSA approaches.

\subsubsection{FTLR Experimental Configuration}
FTLR is a fine‑grained semantic‑matching‑based requirement‑to‑code traceability technique consisting of two phases: precomputation and link processing. In the precomputation phase, requirements and source code are split into sentence‑level and method‑level units respectively. After text preprocessing, fastText\cite{grave2018learning} is adopted to generate embeddings, and the similarity matrix among units is calculated via WMD. In the link‑processing phase, a dual‑threshold mechanism filters valid links: candidate links are first produced by majority voting over method units for each requirement, and valid links are yielded after final‑threshold filtering.

This study extends the FTLR framework to support traceability‑link establishment from use‑case specifications to source code, and enables comparison between two metrics: cosine similarity and word‑mover’s distance.

\subsubsection{LISSA Experimental Configuration}
LISSA is a traceability‑link‑recovery method based on RAG and LLMs. Its workflow mainly comprises three stages: 1) Preprocessing: various software artifacts are structurally decomposed according to artifact types (e.g., splitting requirements into sentences and source code into methods). After textualization, artifact contents are vectorized and stored via an embedding model. 2) Retrieval: for each requirement element, the top‑$k$ most similar target units are retrieved as candidate links according to cosine similarity. 3) Mapping: an LLM performs classification judgment on candidate links, supporting two strategies: plain prompting and Chain‑of‑Thought (CoT) reasoning.

To achieve relatively favorable performance, the Chain‑of‑Thought reasoning strategy is adopted for judgment. Traceability‑link‑recovery experiments are conducted at two granularities: file level and element level. Furthermore, targeting the changed traceable artifacts for the LISSA framework, three aggregation strategies are reconstructed in our experiments, corresponding to Equation (4), Equation (5), and Equations (6)(7). Since LISSA leverages LLMs for binary classification without outputting similarity scores, all three aggregation strategies adopt coverage ratio for aggregation.

Regarding experimental environment configuration, all‑MiniLM‑L6‑v2.onnx is uniformly used for vector embedding, and gpt‑4o‑2024‑05‑13 is employed to perform traceability‑link‑recovery judgments, keeping consistency with the original paper’s experimental settings. Note that LISSA currently only supports Java parsing; hence experiments for the LibEST (C‑language) project are excluded.

Notation Definitions:
\begin{align*}
F&:\text{Code file}\\
R&:\text{Requirement file}\\
U&:\text{Use‑case specification file (belongs to requirement file }R\text{)}\\
E_R&:\text{Set of requirement elements for requirement file }R\\
E_U&:\text{Set of use‑case‑specification elements for use‑case file }U\\
E_F&:\text{Set of code elements for code file }F\\
U(R)&:\text{Set of all use‑case specification files contained in requirement file }R\\
L&\subseteq \biggl(E_R \cup \bigcup_{u \in U(R)} E_u\biggr) \times E_F:\text{Fine‑grained element‑level traceability links}
\end{align*}

\subsubsection{File‑Level Relative‑Coverage Aggregation}
This strategy aggregates judgment results between use‑case‑specification files and code files into requirement‑to‑code‑file traceability links. A traceability link is established between requirement file $R$ and code file $F$ if more than $\theta_1$ fraction of use‑case‑specification files contained in $R$ have traceability links with $F$:
\begin{equation}
\mathrm{Trace}_1(R,F)=\mathds{1}\biggl(
\frac{\bigl|\bigl\{U\in U(R)\,\big|\,\mathrm{Link}(U,F)=1\bigr\}\bigr|}{|U(R)|}
\geq \theta_1
\biggr)
\label{eq:trace1}
\end{equation}

\subsubsection{Dual‑Direction Relative‑Coverage Threshold Judgment}
This strategy directly aggregates judgment results between requirement elements and code elements into requirement‑to‑code‑file traceability links. A traceability link is established between requirement file $R$ and code file $F$ if the fraction of requirement elements within $R$ linked to $F$ reaches $\theta_2$, or the fraction of code elements within $F$ linked to $R$ reaches $\theta_2$:
\begin{equation}
\begin{aligned}
\mathrm{Trace}_2(R,F)=\mathds{1}\biggl(
&\frac{|\{r\in E_R\,|\,\exists s\in E_F,(r,s)\in L\}|}{|E_R|}\geq \theta_2 \\
&\quad \lor \frac{|\{s\in E_F\,|\,\exists r\in E_R,(r,s)\in L\}|}{|E_F|}\geq \theta_2
\biggr)
\end{aligned}
\label{eq:trace2}
\end{equation}

\subsubsection{Fine‑Grained Multi‑Stage Cascade Aggregation}
This strategy performs two‑level aggregation. It first aggregates element‑level judgments between use‑case elements and code elements into use‑case‑specification‑to‑code‑file traceability links, and further aggregates those results into requirement‑to‑code‑file traceability links.

First‑stage aggregation: For use‑case‑specification file $U$ and code file $F$, a traceability link between $U$ and $F$ is established if the fraction of elements in $U$ linked to $F$ reaches $\alpha_1$, or the fraction of elements in $F$ linked to $U$ reaches $\alpha_1$:
\begin{equation}
\begin{aligned}
\mathrm{Link}\,(U,F)=\mathds{1}\biggl(
&\frac{|\{e_u\in E_U\,|\,\exists s\in E_F,(e_u,s)\in L\}|}{|E_U|}\geq \alpha_1 \\
&\quad \lor \frac{|\{s\in E_F\,|\,\exists e_u\in E_U,(e_u,s)\in L\}|}{|E_F|}\geq \alpha_1
\biggr)
\end{aligned}
\label{eq:link_uf}
\end{equation}

Second‑stage aggregation: For requirement file $R$ and code file $F$, a traceability link is established between $R$ and $F$ if the fraction of use‑case specifications within $R$ linked to $F$ reaches $\alpha_2$:
\begin{equation}
\mathrm{Trace}_3(R,F)=\mathds{1}\biggl(\frac{|\{U\in U(R)\,|\,\mathrm{Link}(U,F)=1\}|}{|U(R)|}\geq \alpha_2\biggr)
\label{eq:trace3}
\end{equation}

\subsection{Quality Control}
To guarantee the syntactic‑structural and semantic‑logical compliance of LLM‑generated CNL‑B‑compliant use‑case specifications, three quality‑control strategies are adopted: prompt optimization, automatic format checking, and manual sampling audit. First, prompts are optimized through multi‑round iterations and cross‑validated over three models: DeepSeek‑V3.2, gemini‑3.1‑pro‑preview and Qwen3‑Max‑Instruct, to ensure prompt generalization capability. Second, format normalization and format validation are embedded into the use‑case‑specification generation pipeline to enforce compliance with predefined CNL‑B writing patterns. Moreover, the eANCI dataset is selected as the benchmark. All its generated use‑case specifications undergo manual content inspection and format review to eliminate potential logical deviations and further refine the prompts.

\section{Experimental Results and Analysis}

\subsection{RQ1: To what extent do automatically generated normalized use‑case specifications preserve consistency with the original requirements?}

\begin{table}[htbp]
  \centering
  \caption{Statistics of Randomly Sampled Samples}
  \label{tab:sample_stat}
  \begin{tabular}{lcccc}
    \toprule
    Evaluation Dimension & \multicolumn{2}{c}{Annotator A} & \multicolumn{2}{c}{Annotator B} \\
    \cmidrule(lr){2-3} \cmidrule(lr){4-5}
             & Mean    & Std. Dev.     & Mean    & Std. Dev.     \\
    \midrule
    Accuracy   & 4.480   & 0.193      & 4.488   & 0.175      \\
    Completeness   & 4.477   & 0.187      & 4.423   & 0.178      \\
    Compliance   & 4.768   & 0.423      & 4.815   & 0.338      \\
    \bottomrule
  \end{tabular}
\end{table}

\begin{figure}[htbp]
  \centering
  \includegraphics[width=\linewidth]{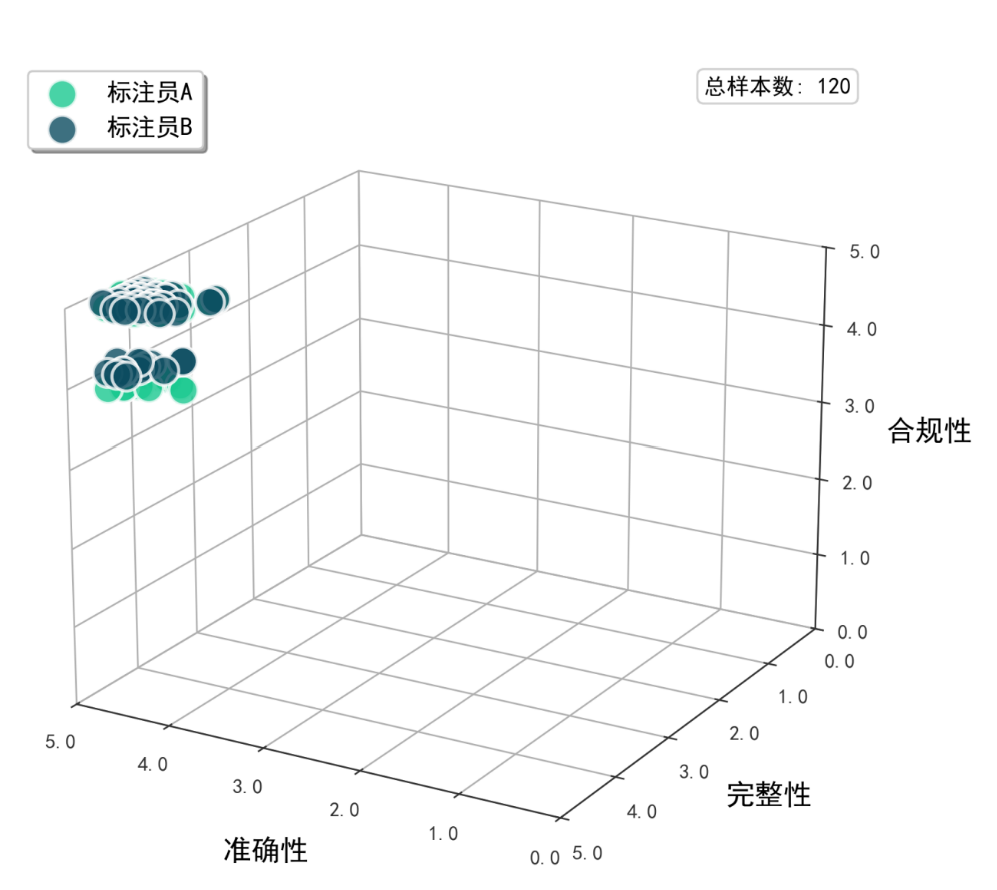}
  \caption{Multi‑dimensional Quantitative Score Distribution for Randomly Sampled Samples}
  \label{fig:fig6}
\end{figure}

Building upon the aforementioned quality‑assurance measures, systematic sampling‑based evaluation is conducted to assess the consistency between generated use‑case specifications and original requirements. For three datasets including eTour, iTrust and LibEST, 60 samples are randomly selected from a total of 321 generated outputs for quality inspection. Two graduate students independent of the authors are invited as separate annotators. They perform 5‑point scoring across three dimensions: accuracy (fidelity to original requirements), completeness (coverage of key logic), and compliance (conformance to CNL‑B format). The inter‑annotator agreement coefficient is computed to evaluate scoring reliability.

Scoring results over the 60 sampled samples show that Annotator A achieves average scores of 4.480, 4.477 and 4.768 for accuracy, completeness and compliance respectively; Annotator B obtains average scores of 4.488, 4.423 and 4.815 (full score = 5). Variances for all dimensions are below 0.5, indicating favorable semantic stability of generated outputs. Scores are concentrated within the range of 4‑5, with no samples scoring lower than 3. The Cohen's Kappa coefficient for independent annotations from the two annotators reaches 0.8323, corresponding to substantial agreement and demonstrating high reliability of scoring results.

\begin{figure*}[htbp]
  \centering
  \includegraphics[width=0.9\textwidth]{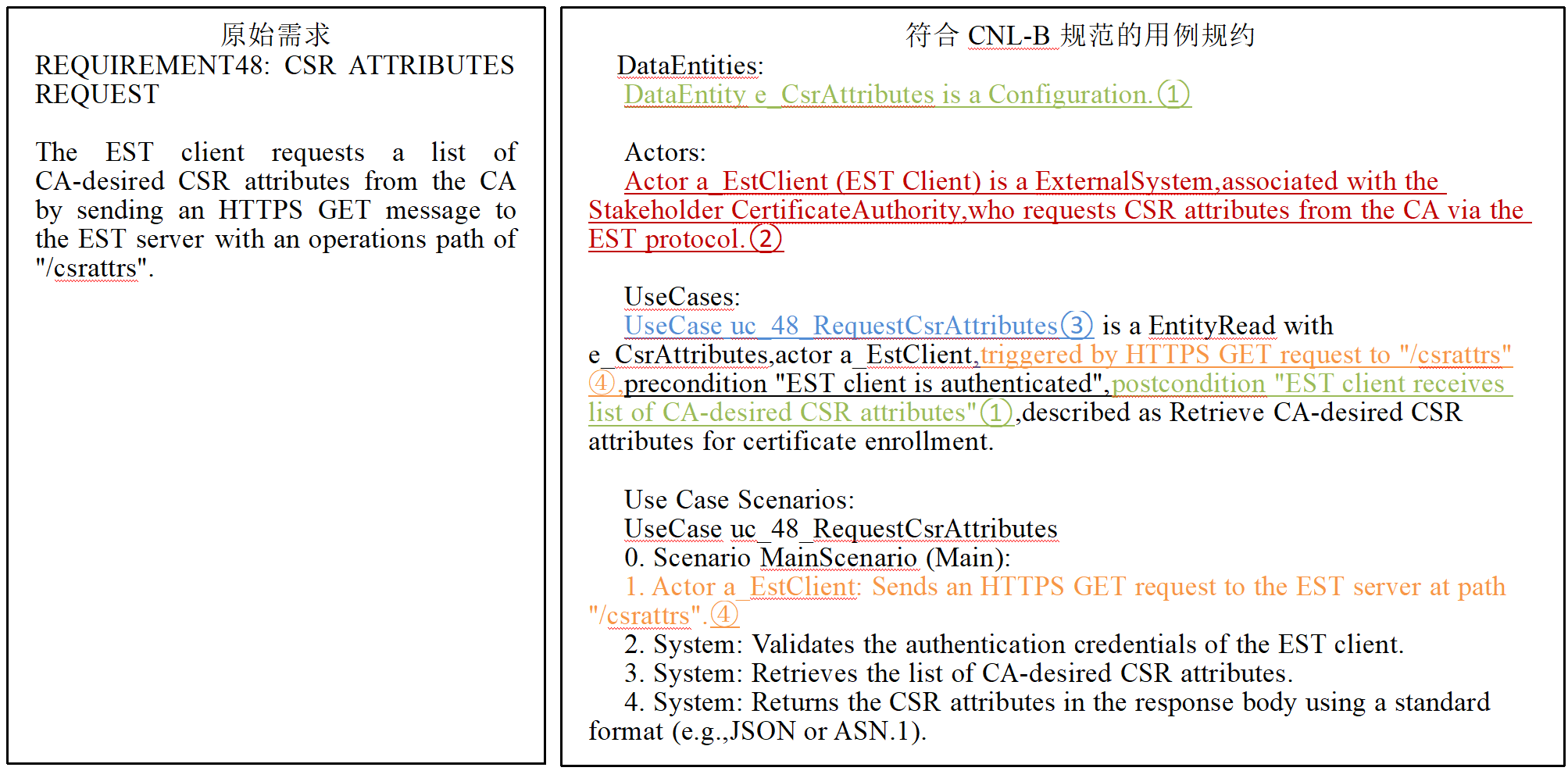}
  \caption{Transformation Process from Original Requirement to CNL‑B‑Compliant Use‑Case Specification}
  \label{fig:fig7}
\end{figure*}

\begin{table*}[htbp]
  \centering
  \caption{Comparison between Original Requirement and Corresponding Use‑Case Specification}
  \label{tab:req_uc_compare}
  \begin{tabular}{p{0.42\textwidth}p{0.20\textwidth}p{0.30\textwidth}}
    \toprule
    \centering\arraybackslash Original Requirement
    & \centering\arraybackslash Corresponding Use‑Case Specification
    & \centering\arraybackslash Correspondence Analysis \\
    \midrule
    EST client requests...CSR attributes
    & \centering \ding{174}
    & The use‑case name directly reflects the core action “request”. \\
    ...from the CA...
    & \centering \ding{173}
    & The requester is explicitly identified as the EST client. \\
    ...list of CA‑desired CSR attributes
    & \centering \ding{172}
    & The core requested data “list of CA‑desired CSR attributes” is clearly stated. \\
    ...by sending an HTTPS GET...path of "/csrattrs"
    & \centering \ding{175}
    & The HTTP method and operation path of the request are precisely matched. \\
    \bottomrule
  \end{tabular}
\end{table*}

Taking REQUIREMENT 48 from the LibEST dataset as an example, Figure \ref{fig:fig7} illustrates its transformation into a CNL‑B‑compliant use‑case specification. As shown in the comparative analysis in Table \ref{tab:req_uc_compare}, core elements from the original requirement are accurately mapped into the generated use‑case specification: the initiator (``EST client'') corresponds to actor a\_EstClient; the requested object (``list of CA‑desired CSR attributes'') corresponds to data entity e\_CsrAttributes; the core operation (``requests'') corresponds to use‑case name uc\_48\_RequestCsrAttributes; and operational details (``HTTPS GET...path of '/csrattrs''' '') correspond to concrete steps within scenarios. Moreover, the generated specification provides reasonable refinements and supplements for the original requirement by clarifying preconditions for execution and internal system processing steps. Such supplementary information conforms to the technical background of the EST protocol and does not exceed the intent of the original requirement.

The above results demonstrate that generated normalized use‑case specifications maintain good consistency with original requirements. Quantitatively, average scores for all three dimensions exceed 4.45, indicating semantic preservation above 89\%. Compliance achieves the highest score, which reveals that LLMs can well respect CNL‑B formatting constraints. Scoring agreement and stability further validate the credibility of generation quality. From the case‑mapping perspective, LLMs can accurately identify core elements within original requirements and map them into structured CNL‑B fields without information loss. Collectively, automatically generated normalized use‑case specifications effectively preserve the semantic logic of original requirements and deliver high‑quality input data for subsequent traceability tasks.

\subsection{RQ2: Compared with original requirements, to what extent can normalized use‑case specifications improve traceability performance (F1‑score) under different traceability frameworks (i.e., IR‑based framework and RAG‑LLM framework)?}

To answer this research question, traceability performance of original requirements and CNL‑B‑compliant use‑case specifications are compared under two representative traceability frameworks, FTLR and LISSA, across different granularities and similarity computation methods. The FTLR framework adopts two metrics: cosine similarity and Word‑Mover’s Distance, and conducts traceability at both file‑level and element‑level granularities. Built upon retrieval‑augmented generation and LLMs, the LISSA framework evaluates the effectiveness of normalized use‑case specifications at file‑level and element‑level granularities with different aggregation strategies.

Under the FTLR framework, normalized use‑case specifications yield performance improvements on most datasets. As shown in Table \ref{tab:tab4_ftlr}, at both file‑level and element‑level granularities, whether cosine similarity (Cos) or Word‑Mover’s Distance (WMD) is adopted for text similarity calculation, replacing original requirements with CNL‑B‑compliant use‑case specifications increases the F1‑score to varying degrees for the iTrust, EanciTrans, and Libset datasets. The most prominent improvement occurs for element‑level traceability using WMD on the Libset dataset, where the F1‑score rises from 0.065 to 0.579, mitigating WMD metric failure caused by overly concise original requirements.

This result indicates that normalization enriches contextual semantic information of requirements and optimizes the representation quality of requirement elements, thus enhancing the recovery capability of traceability links between requirements and source code.

\begin{table*}[htbp]
\centering
\caption{Impact of CNL‑B‑Compliant Use‑Case Specifications on Traceability Performance under the FTLR Framework}
\label{tab:tab4_ftlr}
\resizebox{\textwidth}{!}{%
\begin{tabular}{llcccccccc}
\toprule
Dataset & Metric
& \shortstack{Req‑File‑\\Code‑File (Cos)}
& \shortstack{UC‑File‑\\Code‑File (Cos)}
& \shortstack{Req‑File‑\\Code‑File (WMD)}
& \shortstack{UC‑File‑\\Code‑File (WMD)}
& \shortstack{Req‑Elem‑\\Code‑Method (Cos)}
& \shortstack{UC‑Elem‑\\Code‑Method (Cos)}
& \shortstack{Req‑Elem‑\\Code‑Method (WMD)}
& \shortstack{UC‑Elem‑\\Code‑Method (WMD)} \\
\midrule
\multirow{3}{*}{etour}
& Precision & 0.415 & 0.279 & 0.496 & 0.327 & 0.550 & 0.317 & 0.501 & 0.294 \\
& Recall    & 0.380 & 0.263 & 0.367 & 0.474 & 0.432 & 0.425 & 0.597 & 0.588 \\
& F1‑score  & \textbf{0.397} & 0.271$\downarrow$ (-31.695\%) & \textbf{0.422} & 0.387$\downarrow$ (-8.230\%) & \textbf{0.484} & 0.363$\downarrow$ (-24.864\%) & \textbf{0.545} & 0.392$\downarrow$ (-28.115\%) \\
\midrule
\multirow{3}{*}{Itrust}
& Precision & 0.141 & 0.258 & 0.166 & 0.323 & 0.333 & 0.396 & 0.285 & 0.248 \\
& Recall    & 0.269 & 0.259 & 0.311 & 0.255 & 0.203 & 0.280 & 0.206 & 0.308 \\
& F1‑score  & 0.185 & \textbf{0.258}$\uparrow$ (+39.711\%) & 0.217 & \textbf{0.285}$\uparrow$ (+31.408\%) & 0.252 & \textbf{0.328}$\uparrow$ (+30.017\%) & 0.239 & \textbf{0.275}$\uparrow$ (+14.883\%) \\
\midrule
\multirow{3}{*}{EanciTrans}
& Precision & 0.178 & 0.256 & 0.185 & 0.362 & 0.350 & 0.496 & 0.421 & 0.526 \\
& Recall    & 0.208 & 0.924 & 0.270 & 0.554 & 0.236 & 0.363 & 0.192 & 0.358 \\
& F1‑score  & 0.192 & \textbf{0.401}$\uparrow$ (+109.363\%) & 0.220 & \textbf{0.438}$\uparrow$ (+99.062\%) & 0.282 & \textbf{0.420}$\uparrow$ (+48.722\%) & 0.264 & \textbf{0.426}$\uparrow$ (+61.372\%) \\
\midrule
\multirow{3}{*}{Libset}
& Precision & 0.288 & 0.315 & 0.330 & 0.386 & 0.395 & 0.455 & 0.700 & 0.450 \\
& Recall    & 0.980 & 0.951 & 0.868 & 0.730 & 0.824 & 0.819 & 0.034 & 0.814 \\
& F1‑score  & 0.445 & \textbf{0.473}$\uparrow$ (+6.227\%) & 0.478 & \textbf{0.505}$\uparrow$ (+5.666\%) & 0.534 & \textbf{0.585}$\uparrow$ (+9.551\%) & 0.065 & \underline{\textbf{0.579}}$\uparrow$ (+785.664\%) \\
\bottomrule
\end{tabular}
}
\begin{tablenotes}
\item[Note:] $\uparrow/\downarrow$ denotes F1‑score improvement/degradation of the use‑case‑based approach compared with the original‑requirement‑based approach. \textbf{Bold font} marks the better value within each pair; \underline{underlined font} marks the most prominent improvement. Values in parentheses indicate relative magnitude of change.
\end{tablenotes}
\end{table*}

Under the LISSA framework, the performance of normalized use‑case specifications is strongly coupled with aggregation strategies. Since normalization splits one original requirement into multiple use‑case specifications (i.e., one‑to‑many mapping), directly adopting LISSA’s default “existence‑based aggregation” strategy amplifies noise from random spurious associations. Therefore, relative‑coverage aggregation and multi‑stage cascade aggregation are introduced to mitigate such interference. Experiments demonstrate that combining normalized use‑case specifications with properly designed aggregation strategies yields performance gains. Taking the eAnciTrans dataset (Table \ref{tab:tab5_lissa}) as an example: for file‑level traceability, when paired with relative‑coverage aggregation, the F1‑score increases from 0.212 to 0.324, corresponding to a 52.800\% improvement; for element‑level traceability combined with multi‑stage cascade aggregation, the F1‑score rises from 0.211 to 0.416, achieving a 96.869\% improvement.

\begin{table*}[htbp]
\centering
\caption{Impact of CNL‑B‑Compliant Use‑Case Specifications on Traceability Performance under the LISSA Framework}
\label{tab:tab5_lissa}
\resizebox{\textwidth}{!}{%
\begin{tabular}{llcccccc}
\toprule
Dataset & Metric
& \shortstack{Req‑File‑\\Code‑File\\(existence‑based)}
& \shortstack{UC‑File‑\\Code‑File\\(existence‑based)}
& \shortstack{UC‑File‑\\Code‑File\\(relative‑coverage)}
& \shortstack{Req‑Elem‑\\Code‑Method\\(existence‑based)}
& \shortstack{UC‑Elem‑\\Code‑Method\\(existence‑based)}
& \shortstack{UC‑Elem‑\\Code‑Method\\(multi‑stage cascade)} \\
\midrule
\multirow{3}{*}{EanciTrans}
& Precision & 0.28  & 0.694 & 0.694 & 0.172 & 0.466 & 0.466 \\
& Recall    & 0.171 & 0.212 & 0.212 & 0.273 & 0.376 & 0.376 \\
& F1‑score  & 0.212 & \textbf{0.324}$\uparrow$(+52.800\%) & \textbf{0.324}$\uparrow$(+52.789\%)
           & 0.211 & \underline{\textbf{0.416}}$\uparrow$(+96.869\%) & \underline{\textbf{0.416}}$\uparrow$(+96.869\%) \\
\midrule
\multirow{3}{*}{etour}
& Precision & 0.521 & 0.469 & 0.488 & 0.145 & 0.141 & 0.183 \\
& Recall    & 0.523 & 0.523 & 0.51  & 0.37  & 0.458 & 0.409 \\
& F1‑score  & \textbf{0.522} & 0.495$\downarrow$(-5.223\%) & 0.498$\downarrow$(-4.499\%)
           & 0.208 & 0.216$\uparrow$(+3.717\%) & 0.253$\uparrow$(+21.377\%) \\
\midrule
\multirow{3}{*}{Itrust}
& Precision & 0.08  & 0.047 & 0.049 & 0.229 & 0.071 & 0.317 \\
& Recall    & 0.07  & 0.066 & 0.066 & 0.381 & 0.612 & 0.287 \\
& F1‑score  & \textbf{0.075} & 0.055$\downarrow$(-26.553\%) & 0.057$\downarrow$(-24.431\%)
           & \textbf{0.286} & 0.127$\downarrow$(-55.764\%) & 0.301$\uparrow$(+5.039\%) \\
\midrule
\multirow{3}{*}{Libset}
& Precision & 0.5   & 0.485 & 0.489 & ---   & ---   & --- \\
& Recall    & 0.88  & 0.951 & 0.946 & ---   & ---   & --- \\
& F1‑score  & 0.638 & \textbf{0.642}$\uparrow$(+0.691\%) & \textbf{0.644}$\uparrow$(+1.036\%)
           & ---   & ---   & --- \\
\bottomrule
\end{tabular}
}
\parbox{\linewidth}{\small Note: $\uparrow/\downarrow$ denotes F1‑score improvement/degradation of the use‑case‑based approach compared with the original‑requirement‑based approach. \textbf{Bold font} marks the better value within each pair; \underline{underlined font} marks the most prominent improvement. Values in parentheses indicate relative magnitude of change.}
\end{table*}

\begin{figure*}[htbp]
  \centering
  \includegraphics[width=0.9\linewidth]{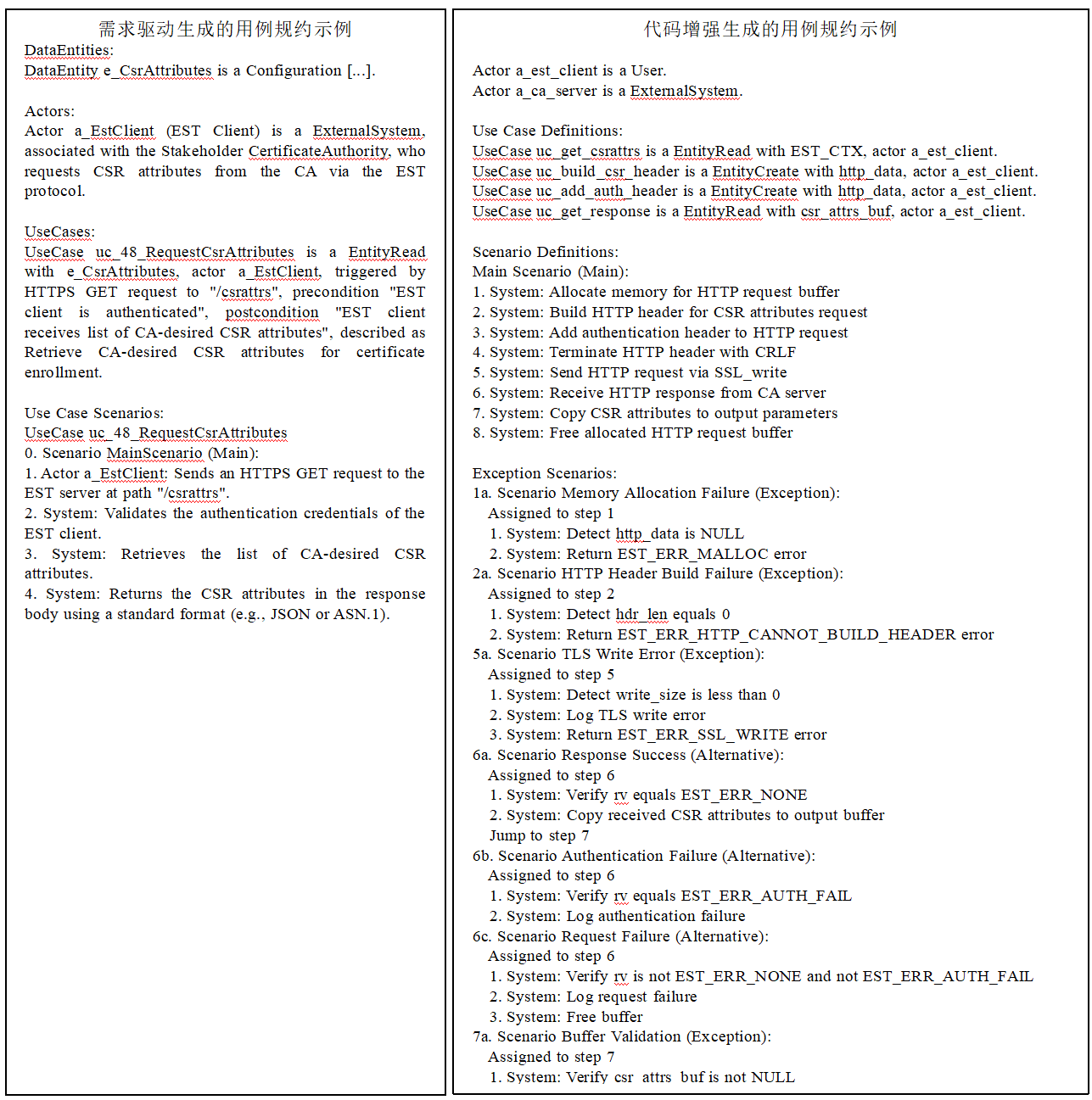}
  \caption{Comparison of Use‑Case Specifications Generated by Requirement‑Driven and Code‑Augmented Approaches}
  \label{fig:fig8}
\end{figure*}

Performance degradation is observed for partial datasets under both frameworks. Under the FTLR framework (Table \ref{tab:tab4_ftlr}), performance drops for the eTour dataset. Under file‑level Word‑Mover’s Distance metric, the F1‑score decreases from 0.422 to 0.387, with a degradation of 8.230\%; under element‑level cosine‑similarity metric, it falls from 0.484 to 0.363, corresponding to a 24.864\% degradation. Under the LISSA framework (Table \ref{tab:tab5_lissa}), for file‑level traceability on eTour with relative‑coverage aggregation, the F1‑score declines from 0.522 to 0.498 (‑4.499\%); for file‑level traceability on iTrust with relative‑coverage aggregation, the F1‑score drops from 0.075 to 0.057. Analysis reveals that original requirements of eTour possess high lexical overlap with source‑code artifacts. When LLMs transform them into lexically constrained normalized use‑case specifications, high‑weight technical terms are converted into low‑weight generic structural words, leading to feature‑word dilution and reduced signal‑to‑noise ratio. For iTrust, original requirements contain limited information and are partially presented in tabular format, which brings challenges for LLM parsing and expansion. Although generated outputs satisfy specification constraints, they exhibit semantic deviations from source code. These two cases conversely verify that requirement quality constitutes a critical factor determining traceability performance.

\begin{table*}[htbp]
\centering
\caption{File‑Level Traceability under FTLR: Comparison between Use‑Case Specifications and Code‑Augmented Variant}
\label{tab:tab6_ftlr_compare}
\resizebox{\textwidth}{!}{%
\begin{tabular}{llcccccc}
\toprule
Dataset & Metric
& Req‑File‑Code‑File (Cos)
& \multicolumn{2}{c}{UC‑File‑Code‑File (Cos)}
& Req‑File‑Code‑File (WMD)
& \multicolumn{2}{c}{UC‑File‑Code‑File (WMD)} \\
\cmidrule(lr){3-3}\cmidrule(lr){4-5}\cmidrule(lr){6-6}\cmidrule(lr){7-8}
& & & Requirement‑Driven & Code‑Augmented & & Requirement‑Driven & Code‑Augmented \\
\midrule
\multirow{3}{*}{etour}
& Precision & 0.415 & 0.279 & 0.268 & 0.496 & 0.327 & 0.311 \\
& Recall    & 0.380 & 0.263 & 0.201 & 0.367 & 0.474 & 0.526 \\
& F1‑score  & \textbf{0.397}
& 0.271$\downarrow$(-31.695\%)
& 0.23$\downarrow$(-42.065\%)
& \textbf{0.422}
& 0.387$\downarrow$(-8.230\%)
& 0.391$\downarrow$(-7.346\%) \\
\midrule
\multirow{3}{*}{Itrust}
& Precision & 0.141 & 0.258 & 0.196 & 0.166 & 0.323 & 0.331 \\
& Recall    & 0.269 & 0.259 & 0.213 & 0.311 & 0.255 & 0.479 \\
& F1‑score  & 0.185
& \textbf{0.258}$\uparrow$(+39.711\%)
& 0.204$\uparrow$(+10.270\%)
& 0.217
& 0.285$\uparrow$(+31.408\%)
& \underline{\textbf{0.391}}$\uparrow$(+80.184\%) \\
\midrule
\multirow{3}{*}{EanciTrans}
& Precision & 0.178 & 0.256 & 0.263 & 0.185 & 0.362 & 0.359 \\
& Recall    & 0.208 & 0.924 & 0.915 & 0.270 & 0.554 & 0.603 \\
& F1‑score  & 0.192
& 0.401$\uparrow$(+109.363\%)
& \underline{\textbf{0.408}}$\uparrow$(+112.500\%)
& 0.220
& 0.438$\uparrow$(+99.062\%)
& \textbf{0.450}$\uparrow$(+104.545\%) \\
\midrule
\multirow{3}{*}{Libset}
& Precision & 0.288 & 0.315 & 0.327 & 0.330 & 0.386 & 0.439 \\
& Recall    & 0.980 & 0.951 & 0.912 & 0.868 & 0.730 & 0.819 \\
& F1‑score  & 0.445
& 0.473$\uparrow$(+6.227\%)
& 0.481$\uparrow$(+8.090\%)
& 0.478
& 0.505$\uparrow$(+5.666\%)
& \underline{\textbf{0.572}}$\uparrow$(+19.665\%) \\
\bottomrule
\end{tabular}
}
\parbox{\linewidth}{\small Note: $\uparrow/\downarrow$ denotes F1‑score improvement/degradation of the use‑case‑based approach compared with the original‑requirement‑based approach. \textbf{Bold font} marks the better value within each pair; \underline{underlined font} marks the most prominent improvement. Values in parentheses indicate relative magnitude of change.}
\end{table*}

\begin{table*}[htbp]
\centering
\caption{Element‑Level Traceability under FTLR: Comparison between Use‑Case Specifications and Code‑Augmented Variant}
\label{tab:tab7_ftlr_element}
\resizebox{\textwidth}{!}{%
\begin{tabular}{llcccccc}
\toprule
Dataset & Metric
& Req‑Elem‑Code‑Method (Cos)
& \multicolumn{2}{c}{UC‑Elem‑Code‑Method (Cos)}
& Req‑Elem‑Code‑Method (WMD)
& \multicolumn{2}{c}{UC‑Elem‑Code‑Method (WMD)} \\
\cmidrule(lr){3-3}\cmidrule(lr){4-5}\cmidrule(lr){6-6}\cmidrule(lr){7-8}
& & & Requirement‑Driven & Code‑Augmented & & Requirement‑Driven & Code‑Augmented \\
\midrule
\multirow{3}{*}{etour}
& Precision & 0.550 & 0.317 & 0.252 & 0.501 & 0.294 & 0.238 \\
& Recall    & 0.432 & 0.425 & 0.594 & 0.597 & 0.588 & 0.714 \\
& F1‑score  & \textbf{0.484}
& 0.363$\downarrow$(-24.864\%)
& 0.354$\downarrow$(-26.796\%)
& \textbf{0.545}
& 0.392$\downarrow$(-28.115\%)
& 0.357$\downarrow$(-34.469\%) \\
\midrule
\multirow{3}{*}{Itrust}
& Precision & 0.333 & 0.396 & 0.305 & 0.285 & 0.248 & 0.259 \\
& Recall    & 0.203 & 0.280 & 0.402 & 0.206 & 0.308 & 0.392 \\
& F1‑score  & 0.252
& 0.328$\uparrow$(+30.017\%)
& \textbf{0.347}$\uparrow$(+37.662\%)
& 0.239
& 0.275$\uparrow$(+14.883\%)
& \underline{\textbf{0.312}}$\uparrow$(+30.353\%) \\
\midrule
\multirow{3}{*}{EanciTrans}
& Precision & 0.350 & 0.496 & 0.608 & 0.421 & 0.526 & 0.529 \\
& Recall    & 0.236 & 0.363 & 0.517 & 0.192 & 0.358 & 0.624 \\
& F1‑score  & 0.282
& 0.420$\uparrow$(+48.722\%)
& 0.559$\uparrow$(+98.095\%)
& 0.264
& 0.426$\uparrow$(+61.372\%)
& \underline{\textbf{0.573}}$\uparrow$(+116.976\%) \\
\midrule
\multirow{3}{*}{Libset}
& Precision & 0.395 & 0.455 & 0.445 & 0.700 & 0.450 & 0.445 \\
& Recall    & 0.824 & 0.819 & 0.912 & 0.034 & 0.814 & 0.877 \\
& F1‑score  & 0.534
& 0.585$\uparrow$(+21.973\%)
& 0.598$\uparrow$(+11.985\%)
& 0.065
& 0.579$\uparrow$(+785.664\%)
& \underline{\textbf{0.591}}$\uparrow$(+808.860\%) \\
\bottomrule
\end{tabular}
}
\parbox{\linewidth}{\small Note: $\uparrow/\downarrow$ denotes F1‑score improvement/degradation of the use‑case‑based approach compared with the original‑requirement‑based approach. \textbf{Bold font} marks the better value within each pair; \underline{underlined font} marks the most prominent improvement. Values in parentheses indicate relative magnitude of change.}
\end{table*}

\begin{table*}[htbp]
\centering
\caption{File‑Level Traceability under LISSA: Comparison between Use‑Case Specifications and Code‑Augmented Variant}
\label{tab:tab8_lissa_file}
\resizebox{\textwidth}{!}{%
\begin{tabular}{llccccc}
\toprule
Dataset & Metric
& \shortstack{Req‑File‑Code‑File\\(existence‑based aggregation)}
& \multicolumn{2}{c}{\shortstack{UC‑File‑Code‑File\\(existence‑based aggregation)}}
& \multicolumn{2}{c}{\shortstack{UC‑File‑Code‑File\\(relative‑coverage aggregation)}} \\
\cmidrule(lr){3-3}\cmidrule(lr){4-5}\cmidrule(lr){6-7}
& & & Requirement‑Driven & Code‑Augmented & Requirement‑Driven & Code‑Augmented \\
\midrule
\multirow{3}{*}{EanciTrans}
& Precision & 0.280 & 0.694 & 0.511 & 0.694 & 0.528 \\
& Recall    & 0.171 & 0.212 & 0.511 & 0.212 & 0.504 \\
& F1‑score  & 0.212
& 0.324$\uparrow$(+52.800\%)
& 0.511$\uparrow$(+140.756\%)
& 0.324$\uparrow$(+52.789\%)
& \underline{\textbf{0.516}}$\uparrow$(+143.011\%) \\
\midrule
\multirow{3}{*}{etour}
& Precision & 0.521 & 0.469 & 0.183 & 0.488 & 0.441 \\
& Recall    & 0.523 & 0.523 & 0.727 & 0.510 & 0.471 \\
& F1‑score  & \textbf{0.522}
& 0.495$\downarrow$(-5.223\%)
& 0.293$\downarrow$(-43.930\%)
& 0.498$\downarrow$(-4.499\%)
& 0.455$\downarrow$(-12.758\%) \\
\midrule
\multirow{3}{*}{Itrust}
& Precision & 0.080 & 0.047 & 0.022 & 0.049 & 0.059 \\
& Recall    & 0.070 & 0.066 & 0.259 & 0.066 & 0.192 \\
& F1‑score  & \textbf{0.075}
& 0.055$\downarrow$(-26.553\%)
& 0.041$\downarrow$(-45.438\%)
& 0.057$\downarrow$(-24.431\%)
& 0.091$\uparrow$(+21.178\%) \\
\midrule
\multirow{3}{*}{Libset}
& Precision & 0.500 & 0.485 & 0.464 & 0.489 & 0.478 \\
& Recall    & 0.880 & 0.951 & 0.995 & 0.946 & 0.989 \\
& F1‑score  & 0.638
& 0.642$\uparrow$(+0.691\%)
& 0.633$\downarrow$(-0.718\%)
& 0.644$\uparrow$(+1.036\%)
& 0.644$\uparrow$(+1.004\%) \\
\bottomrule
\end{tabular}
}
\parbox{\linewidth}{\small Note: $\uparrow/\downarrow$ denotes F1‑score improvement/degradation of the use‑case‑based approach compared with the original‑requirement‑based approach. \textbf{Bold font} marks the better value within each pair; \underline{underlined font} marks the most prominent improvement. Values in parentheses indicate relative magnitude of change.}
\end{table*}

\begin{table*}[htbp]
\centering
\caption{Element‑Level Traceability under LISSA: Comparison between Use‑Case Specifications and Code‑Augmented Variant}
\label{tab:tab9_lissa_element}
\resizebox{\textwidth}{!}{%
\begin{tabular}{llccccc}
\toprule
Dataset & Metric
& \shortstack{Req‑Elem‑Code‑Method\\(existence‑based aggregation)}
& \multicolumn{2}{c}{\shortstack{UC‑Elem‑Code‑Method\\(existence‑based aggregation)}}
& \multicolumn{2}{c}{\shortstack{UC‑Elem‑Code‑Method\\(multi‑stage cascade aggregation)}} \\
\cmidrule(lr){3-3}\cmidrule(lr){4-5}\cmidrule(lr){6-7}
& & & Requirement‑Driven & Code‑Augmented & Requirement‑Driven & Code‑Augmented \\
\midrule
\multirow{3}{*}{EanciTrans}
& Precision & 0.172 & 0.466 & 0.341 & 0.466 & 0.341 \\
& Recall    & 0.273 & 0.376 & 0.743 & 0.376 & 0.743 \\
& F1‑score  & 0.211
& 0.416$\uparrow$(+96.869\%)
& \underline{\textbf{0.468}}$\uparrow$(+121.365\%)
& 0.416$\uparrow$(+96.869\%)
& 0.468$\uparrow$(+121.233\%) \\
\midrule
\multirow{3}{*}{etour}
& Precision & 0.145 & 0.141 & 0.088 & 0.183 & 0.273 \\
& Recall    & 0.370 & 0.458 & 0.519 & 0.409 & 0.231 \\
& F1‑score  & 0.208
& 0.216$\uparrow$(+3.717\%)
& 0.151$\downarrow$(-27.612\%)
& 0.253$\uparrow$(+21.377\%)
& 0.250$\uparrow$(+20.175\%) \\
\midrule
\multirow{3}{*}{Itrust}
& Precision & 0.229 & 0.071 & 0.045 & 0.317 & 0.335 \\
& Recall    & 0.381 & 0.612 & 0.927 & 0.287 & 0.409 \\
& F1‑score  & \textbf{0.286}
& 0.127$\downarrow$(-55.764\%)
& 0.086$\downarrow$(-70.048\%)
& 0.301$\uparrow$(+5.039\%)
& \underline{\textbf{0.369}}$\uparrow$(+28.637\%) \\
\bottomrule
\end{tabular}
}
\parbox{\linewidth}{\small Note: $\uparrow/\downarrow$ denotes F1‑score improvement/degradation of the use‑case‑based approach compared with the original‑requirement‑based approach. \textbf{Bold font} marks the better value within each pair; \underline{underlined font} marks the most prominent improvement. Values in parentheses indicate relative magnitude of change.}
\end{table*}

To further explore the effectiveness boundary of normalizing requirements into use‑case specifications, this study introduces a code‑augmented approach (see Appendix B for details) as a comparison baseline equipped with prior knowledge. Under the FTLR and LISSA frameworks (as shown in Tables \ref{tab:tab6_ftlr_compare} to \ref{tab:tab9_lissa_element}), we observe that the code‑augmented approach achieves higher traceability performance on most datasets by supplementing code‑derived information. Nevertheless, in certain scenarios (e.g., the FTLR cosine metric for eTour and iTrust), the requirement‑driven approach outperforms its code‑augmented counterpart.

To explain this counter‑intuitive observation, our analysis reveals that although code details are generally beneficial, excessive injection of low‑level code information introduces semantic interference toward high‑level requirements. This interference manifests in two aspects. First, the introduction of technical vocabulary increases false positives. When extracting code summaries, the code‑augmented approach inevitably imports numerous generic technical terms (e.g., Manager, Context, List). In cosine‑similarity retrieval based on lexical co‑occurrence, these low‑discriminative terms dilute core business features and produce artificially inflated similarity scores between use‑case specifications and many irrelevant code files. For instance, the precision drops from 0.258 to 0.196 for the iTrust dataset in Table \ref{tab:tab6_ftlr_compare}, which stems exactly from increased false positives caused by lexical noise.

Second, the coverage‑check mechanism embedded in the code‑augmented approach (Algorithm \ref{alg:iterative_decomposition_2}) enforces use‑case specifications to cover and associate all linked code files. Real‑world traceability links often span multiple architectural layers. Forcing heterogeneous code logic into a single normalized use‑case violates the single‑responsibility principle of use cases. Such semantic overload causes misjudgments during matching due to shifted or unfocused semantics. As illustrated in Figure \ref{fig:fig8}, use‑case specifications generated by the code‑augmented approach contain far more implementation‑oriented technical details for the same original requirement.

In summary, the experimental comparisons verify the double‑edged‑sword effect of code context when bridging the semantic gap. Meanwhile, they highlight the core advantage of the proposed requirement‑driven approach: without extra computational overhead from code processing or prior ground‑truth knowledge, the requirement‑driven method maintains pure business logic and a proper level of abstraction, yielding high cost‑effectiveness. It achieves near‑upper‑bound performance in most scenarios while effectively avoiding noise from low‑level implementation details, which delivers stronger robustness and practical value for real‑world industrial applications.

From the above experimental results, the following conclusions can be drawn. Normalized use‑case specifications improve traceability performance across most datasets and traceability frameworks, and the improvement is particularly prominent when original requirements are of low quality. The choice of aggregation strategy exerts a considerable impact on traceability performance; appropriate aggregation helps filter noise and boost precision. Requirement‑driven normalized use‑cases achieve performance close to that of the code‑augmented variant under most conditions, demonstrating the high cost‑effectiveness of the generation method proposed in this work. Meanwhile, normalization may bring slight performance degradation for requirements with high lexical overlap with source code. This suggests that practitioners should decide whether to apply normalization according to the quality of input requirements in practical deployment.

\subsection{RQ3: What are the differences of different requirement‑splitting strategies on generated use‑case specifications and traceability performance?}

\begin{table*}[htbp]
  \centering
  \caption{Comparison of the number of use‑cases generated by different splitting strategies}
  \label{tab:tab10_case_num}
  \begin{tabular}{lccc}
    \toprule
    Dataset & Number of traced requirement files & Number of use‑case specifications under business‑oriented splitting & Number of use‑case specifications under technical‑layer‑oriented splitting \\
    \midrule
    EanciTrans & 39 & 43 & 119 \\
    Libset     & 47 & 71 & 147 \\
    \bottomrule
  \end{tabular}
\end{table*}

\begin{table*}[htbp]
  \centering
  \caption{Performance comparison of different splitting strategies under file‑level cosine similarity}
  \label{tab:tab11_split_perf}
  \resizebox{\textwidth}{!}{%
  \begin{tabular}{llcccc}
    \toprule
    Dataset & Metric
    & \multicolumn{2}{c}{FTLR}
    & \multicolumn{2}{c}{LISSA} \\
    \cmidrule(lr){3-4}\cmidrule(lr){5-6}
    & & Business‑logic‑oriented & Technical‑layer‑oriented & Business‑logic‑oriented & Technical‑layer‑oriented \\
    \midrule
    \multirow{3}{*}{EanciTrans}
    & Precision & 0.256 & 0.257 & 0.694 & 0.657 \\
    & Recall    & 0.924 & 0.931 & 0.212 & 0.243 \\
    & F1‑score  & 0.401 & \textbf{0.403} & 0.324 & \textbf{0.355} \\
    \midrule
    \multirow{3}{*}{Libset}
    & Precision & 0.315 & 0.315 & 0.489 & 0.511 \\
    & Recall    & 0.951 & 0.995 & 0.946 & 0.913 \\
    & F1‑score  & 0.473 & \textbf{0.478} & 0.644 & \textbf{0.655} \\
    \bottomrule
  \end{tabular}
  }
  \parbox{\linewidth}{\small Note: Bold font denotes the better value within each pair of methods.}
\end{table*}

Original requirement documents contain requirements of diverse granularities. Some requirements have single responsibilities and are easy to comprehend and trace, while others mix multiple functions and increase the difficulty of subsequent processing. To generate high‑quality normalized use‑case specifications, this study invokes an LLM to judge and split raw requirements in the preprocessing stage: a requirement item will be split if it covers multiple responsibilities. Practical observations reveal two distinct splitting tendencies produced by the LLM. One is business‑logic‑oriented splitting, which divides requirements according to complete business flows of user operations. The other is technical‑layer‑oriented splitting, which partitions requirements from technical‑implementation perspectives including front‑end interaction, back‑end processing, and data manipulation. These two splitting approaches exert different influences on the quantity of generated use‑case specifications and subsequent traceability performance.

To compare the differences between the two splitting strategies, this study designs two distinct prompts to guide the LLM to perform business‑logic‑oriented and technical‑layer‑oriented splitting on the EanciTrans and Libset datasets, which represent different domains and programming languages respectively. We compare the traceability performance of use‑case specifications generated by the two splitting strategies under file‑level evaluation with cosine similarity for both the FTLR and LISSA traceability frameworks.

In terms of the number of generated use‑cases (Table \ref{tab:tab10_case_num}), technical‑layer‑oriented splitting produces significantly more use‑cases than business‑logic‑oriented splitting. For the EanciTrans dataset, 43 use‑cases are generated by business‑logic‑oriented splitting and 119 use‑cases by technical‑layer‑oriented splitting. For the Libset dataset, 71 use‑cases are generated by business‑logic‑oriented splitting and 147 use‑cases by technical‑layer‑oriented splitting. The number of use‑cases generated by technical‑layer‑oriented splitting is approximately two to three times that of business‑logic‑oriented splitting.

In terms of traceability performance (Table \ref{tab:tab11_split_perf}), under the FTLR framework, the F1‑score of business‑logic‑oriented splitting is 0.401 and that of technical‑layer‑oriented splitting is 0.403 for EanciTrans; for Libset, the F1‑score is 0.473 for business‑logic‑oriented splitting and 0.478 for technical‑layer‑oriented splitting. Under the LISSA framework, the F1‑score of business‑logic‑oriented splitting is 0.324 and that of technical‑layer‑oriented splitting is 0.355 for EanciTrans; for Libset, the F1‑score is 0.644 for business‑logic‑oriented splitting and 0.655 for technical‑layer‑oriented splitting. Overall, technical‑layer‑oriented splitting yields slightly higher F1‑scores, yet the improvement is limited, with differences below 0.02 across both datasets.

Apart from quantitative metrics, use‑case specifications generated by the two splitting strategies differ in readability. Business‑logic‑oriented splitting preserves complete user‑operation workflows. Its outputs conform better to human reading habits and facilitate subsequent manual comprehension and maintenance. By contrast, technical‑layer‑oriented splitting decomposes one business workflow into multiple technical‑implementation steps. Although it establishes clearer correspondences with technical‑implementation layers, it breaks business integrity and increases the complexity of traceability links.

In summary, business‑logic‑oriented splitting achieves traceability performance comparable to technical‑layer‑oriented splitting, while generating fewer use‑cases with lower computational overhead. Meanwhile, its outputs possess better readability for manual comprehension and maintenance. Therefore, business‑logic‑oriented splitting is more feasible for generating normalized use‑case specifications.
\section{Analysis of Factors Affecting Validity}

Although this study adopts the high‑performance gpt‑4o‑2024‑05‑13 model for traceability link judgment in LISSA, large language models are essentially inference systems that perform deep statistical modeling over massive corpora and conduct probabilistic mapping based on high‑dimensional vector representations and context‑dependent relationships. Lacking genuine semantic understanding, such models still suffer from output uncertainty and potential hallucinations when handling highly specialized software‑engineering texts. To mitigate the impact of this factor on experimental results, a rigorous manual quality‑check mechanism is introduced. The Cohen's Kappa coefficient among human reviewers is measured at 0.8323, reaching the standard of “high agreement”, which eliminates systematic errors caused by model output bias as much as possible.

While this study covers projects written in Java and C across four different domains, traceability links in industrial‑scale ultra‑large code repositories can be far more complex. Future work is required to validate the scalability of the proposed approach under ten‑million‑line code volumes.

Experimental results reveal that traceability performance (especially F1‑score) is strongly coupled with the choice of aggregation algorithm. Particularly for fine‑grained element‑level matching, the noise‑filtering capability of the aggregation strategy directly determines final outcomes. Aggregation algorithms adopted in this work, such as bidirectional relative‑coverage‑threshold‑based decision‑making and mode‑mean aggregation, are relatively reasonable solutions determined according to the characteristics of current datasets. Large‑scale horizontal ablation experiments over all plausible aggregation algorithms have not been conducted. Accordingly, the performance conclusions drawn in this paper are conditional to some extent, and future work will target more general‑purpose aggregation paradigms.

\section{Discussion}

\subsection{Applicable Boundaries of Normalized Use‑Case Specifications}
This study finds that normalized use‑case specifications do not consistently yield positive improvements on traceability performance across all scenarios. The anomalous behavior observed on the eTour dataset delineates the applicable boundaries of our proposed method. This dataset exhibits high lexical overlap between original requirements and low‑level source code. Under such circumstances, normalization converts these high‑weight technical terms into generic structural words with low weights, which dilutes discriminative features and degrades signal‑to‑noise ratio. This observation implies that the effectiveness of normalized use‑case specifications is negatively correlated with original‑requirement quality: higher‑quality original requirements leave smaller room for normalization‑driven performance gains; normalization may produce counter‑productive effects when original requirements already align well with source code.

Motivated by this finding, we recommend pre‑evaluation of requirements before applying normalization transformation. For instance, lexical overlap between raw requirements and codebases (e.g., average TF‑IDF cosine similarity) can quantify requirement‑code alignment. If the alignment is already high (e.g., eTour), raw requirements should be retained or only lightweight format normalization shall be applied. When pre‑evaluation indicates vague requirement descriptions and a large semantic gap (e.g., iTrust, LibEST), full CNL‑B normalization transformation can deliver tangible performance improvements.

\subsection{Bottleneck Analysis and Optimization Directions for Traceability Performance}
Experimental results demonstrate that normalized use‑case specifications improve the baseline of traceability performance, reduce occurrences of extremely low‑value samples, and stabilize overall performance. This validates the value of standardizing representations from the requirement side for narrowing the semantic gap. Nevertheless, state‑of‑the‑art performance still has a noticeable gap before practical engineering deployment. Although normalization alleviates semantic ambiguity in requirements, it cannot break the inherent performance upper bound of traceability techniques. Improved input quality guarantees the lower bound of performance, whereas the upper bound remains determined by the matching capability of traceability algorithms.

The underlying reason is that bridging the semantic gap requires collaboration between requirement representations and traceability algorithms. Normalized use‑case specifications resolve the clarity of requirement descriptions, yet understanding code intent still depends on the intrinsic semantic‑matching capacity of traceability algorithms. Traditional lexical‑co‑occurrence‑based frameworks are constrained by shallow lexical overlap. For large‑language‑model‑based frameworks, models still lack sufficient ability to distinguish core information from secondary content within specifications. Therefore, future research shall shift from solely optimizing inputs toward joint optimization of inputs and algorithms. On one hand, key‑information filtering mechanisms can be introduced during specification generation to guide traceability algorithms to focus on core semantic units. On the other hand, efficient traceability techniques can be explored upon normalized use‑case specifications, attempting to break through current performance ceilings.

\subsection{Benefits and Challenges of Code Context (Code‑Augmentation)}
Experiments show that use‑case specifications generated with code context (code‑augmentation) generally outperform counterparts produced purely from requirements, verifying the auxiliary effect of code information for requirement comprehension. However, code‑augmentation yields slightly inferior results in a small number of cases. Analysis attributes this phenomenon to how code information is organized: injecting code snippets with low relevance to target requirements may distract models from core semantics. This finding suggests that introducing code context shall be accompanied by effective information‑filtering strategies rather than simple information stacking. Prioritizing high‑value code information within limited context windows constitutes a key factor for improving generation quality.

\section{Conclusion}
A semantic gap arises between requirements and source code due to divergent abstraction levels. Existing studies mostly focus on algorithm optimization while overlooking quality defects within requirement descriptions themselves. To bridge this gap at the source, this work proposes reconstructing raw requirements automatically into CNL‑B‑compliant use‑case specifications using large language models, so as to unify text structures, filter noise, and supplement implicit domain context.

To systematically evaluate the effectiveness of such normalized use‑case specifications, empirical experiments are conducted on four public datasets over two representative traceability frameworks: information‑retrieval‑based FTLR and retrieval‑augmented‑generation‑based LISSA. Experimental results indicate that normalized use‑case specifications effectively boost requirement‑to‑code traceability performance for requirements with ambiguous original descriptions. The main contributions are summarized as follows:
\begin{enumerate}
    \item A triple‑form dataset of “raw requirement‑normalized use‑case‑code” is constructed. Based on four public datasets, LLM‑driven automatic generation produces CNL‑B‑compliant use‑case specifications. Multi‑round quality evaluations verify the accuracy and reliability of generated data, providing data support for subsequent research within the requirement traceability community.
    \item An automatic use‑case‑specification generation framework is proposed. The pipeline integrates requirement splitting, CNL‑B structured generation, and quality validation, offering a reproducible technical solution for automatic construction of normalized use‑case specifications.
    \item The performance improvement brought by normalized use‑case specifications is quantitatively validated. Experiments show that introducing normalized use‑case specifications yields consistent performance gains across most datasets. For the LibEST dataset, the F1‑score rises from 0.065 to 0.579. This result confirms the effectiveness of narrowing the semantic gap from the requirement side.
    \item Advantages of the business‑logic‑oriented splitting strategy are revealed. Comparisons between business‑logic‑oriented and technical‑layer‑oriented splitting show that the former reduces the volume of normalized use‑cases by 58.6\% on average. Organized around functional‑semantic units, it preserves complete business logic, avoids semantic fragmentation, and achieves stable traceability performance with lower manual and computational costs. This finding delivers practical references for generating normalized use‑case specifications.
\end{enumerate}

\bibliographystyle{IEEEtran}
\bibliography{refer}
\appendix
\subsection*{Appendix A Related Data Tables}
\addcontentsline{toc}{subsection}{Appendix A Related Data Tables}
\vspace{1em}
\mbox{}
\begin{table*}[t]
\centering
\caption{Multi‑dimensional Quality Quantification Scores for Sampled Instances}
\label{tab:12}
\resizebox{\textwidth}{!}{
\begin{tabular}{c c l ccc ccc}
\toprule
No. & Dataset & Use‑case Name
& \multicolumn{3}{c}{Annotator A}
& \multicolumn{3}{c}{Annotator B} \\
\cmidrule(lr){4-6}\cmidrule(lr){7-9}
& &
& \makecell{Accuracy\\(Full score: 5)}
& \makecell{Completeness\\(Full score: 5)}
& \makecell{Compliance\\(Full score: 5)}
& \makecell{Accuracy\\(Full score: 5)}
& \makecell{Completeness\\(Full score: 5)}
& \makecell{Compliance\\(Full score: 5)} \\
\midrule
1 & eTour & UC1\_FGUC\_001.txt & 4.7 & 4.6 & 5 & 4.5 & 4.7 & 5 \\
2 & eTour & UC4\_FGUC\_001.txt & 4.4 & 4.8 & 5 & 4.4 & 4.5 & 5 \\
3 & eTour & UC6\_FGUC\_001.txt & 4.3 & 4.5 & 5 & 4.1 & 4.6 & 5 \\
4 & eTour & UC9\_FGUC\_001.txt & 4.7 & 4.8 & 5 & 4.5 & 4.3 & 5 \\
5 & eTour & UC12\_FGUC\_001.txt & 4.4 & 4.7 & 5 & 4.6 & 4.3 & 5 \\
6 & eTour & UC13\_FGUC\_002.txt & 4.2 & 4.7 & 5 & 4.6 & 4.5 & 5 \\
7 & eTour & UC15\_FGUC\_001.txt & 4.4 & 4.6 & 5 & 4.4 & 4.2 & 5 \\
8 & eTour & UC19\_FGUC\_001.txt & 4.3 & 4.5 & 5 & 4.6 & 4.4 & 5 \\
9 & eTour & UC20\_FGUC\_002.txt & 4.6 & 4.7 & 4 & 4.5 & 4.8 & 4.2 \\
10 & eTour & UC23\_FGUC\_002.txt & 4.6 & 4.5 & 5 & 4.8 & 4.7 & 5 \\
11 & eTour & UC28\_FGUC\_001.txt & 4.8 & 4.4 & 5 & 4.6 & 4.7 & 5 \\
12 & eTour & UC34\_FGUC\_001.txt & 4.6 & 4.3 & 5 & 4.2 & 4.4 & 5 \\
13 & eTour & UC38\_FGUC\_001.txt & 4.1 & 4.5 & 4 & 4.8 & 4.5 & 4.2 \\
14 & eTour & UC41\_FGUC\_001.txt & 4.5 & 4.2 & 5 & 4.7 & 4.6 & 5 \\
15 & eTour & UC49\_FGUC\_001.txt & 4.4 & 4.3 & 5 & 4.4 & 4.5 & 5 \\
16 & eTour & UC52\_FGUC\_001.txt & 4.2 & 4.4 & 5 & 4.6 & 4.5 & 5 \\
17 & eTour & UC53\_FGUC\_002.txt & 4.5 & 4.3 & 4 & 4.3 & 4.2 & 4.2 \\
18 & eTour & UC55\_FGUC\_001.txt & 4.7 & 4.5 & 5 & 4.5 & 4.5 & 5 \\
19 & eTour & UC57\_FGUC\_001.txt & 4.6 & 4.8 & 5 & 4.7 & 4.2 & 5 \\
20 & eTour & UC58\_FGUC\_001.txt & 4.7 & 4.2 & 5 & 4.6 & 4.5 & 5 \\
21 & iTrust & UC1E1\_FGUC\_001.txt & 4.5 & 4.7 & 5 & 4.6 & 4.8 & 5 \\
22 & iTrust & UC3S3\_FGUC\_001.txt & 4.7 & 4.3 & 5 & 4.7 & 4.4 & 5 \\
23 & iTrust & UC3S5\_FGUC\_001.txt & 4.5 & 4.3 & 5 & 4.5 & 4.6 & 5 \\
24 & iTrust & UC4S3\_FGUC\_001.txt & 4.1 & 4.5 & 5 & 4.2 & 4.7 & 5 \\
25 & iTrust & UC6S2\_FGUC\_001.txt & 4.3 & 4.2 & 5 & 4.4 & 4.6 & 5 \\
26 & iTrust & UC11S1\_FGUC\_002.txt & 4.5 & 4.6 & 4 & 4.3 & 4.5 & 4.2 \\
27 & iTrust & UC11S1\_FGUC\_005.txt & 4.2 & 4.4 & 4 & 4.6 & 4.7 & 4.2 \\
28 & iTrust & UC15E1\_FGUC\_001.txt & 4.5 & 4.2 & 5 & 4.6 & 4.4 & 5 \\
29 & iTrust & UC17S2\_FGUC\_001.txt & 4.4 & 4.3 & 4 & 4.5 & 4.6 & 4.2 \\
30 & iTrust & UC21S1\_FGUC\_001.txt & 4.5 & 4.5 & 5 & 4.7 & 4.5 & 5 \\
31 & iTrust & UC27S2\_FGUC\_001.txt & 4.7 & 4.6 & 5 & 4.4 & 4.3 & 5 \\
32 & iTrust & UC30E1\_FGUC\_001.txt & 4.4 & 4.5 & 5 & 4.4 & 4.5 & 5 \\
33 & iTrust & UC30S2\_FGUC\_001.txt & 4.6 & 4.3 & 4 & 4.7 & 4.8 & 4.2 \\
34 & iTrust & UC34E1\_FGUC\_001.txt & 4.5 & 4.6 & 5 & 4.5 & 4.3 & 5 \\
35 & iTrust & UC34S5\_FGUC\_001.txt & 4.6 & 4.7 & 4 & 4.3 & 4.2 & 4.2 \\
36 & iTrust & UC35S1\_FGUC\_002.txt & 4.4 & 4.6 & 4 & 4.5 & 4.5 & 4.2 \\
37 & iTrust & UC37S1\_FGUC\_002.txt & 4.7 & 4.5 & 4 & 4.6 & 4.4 & 4.2 \\
38 & iTrust & UC37S3\_FGUC\_001.txt & 4.3 & 4.4 & 5 & 4.5 & 4.8 & 5 \\
39 & iTrust & UC38S1\_FGUC\_001.txt & 4.5 & 4.7 & 5 & 4.7 & 4.3 & 5 \\
40 & iTrust & UC38S2\_FGUC\_001.txt & 4.6 & 4.3 & 4 & 4.5 & 4.4 & 4.2 \\
41 & LibEST & RQ4\_FGUC\_001.txt & 4.5 & 4.7 & 5 & 4.4 & 4.7 & 5 \\
42 & LibEST & RQ5\_FGUC\_001.txt & 4.3 & 4.6 & 5 & 4.4 & 4.5 & 5 \\
43 & LibEST & RQ8\_FGUC\_001.txt & 4.7 & 4.5 & 5 & 4.3 & 4.6 & 5 \\
44 & LibEST & RQ14\_FGUC\_001.txt & 4.4 & 4.2 & 5 & 4.6 & 4.5 & 5 \\
45 & LibEST & RQ14\_FGUC\_003.txt & 4.3 & 4.5 & 5 & 4.5 & 4.6 & 5 \\
46 & LibEST & RQ16\_FGUC\_002.txt & 4.7 & 4.8 & 4 & 4.6 & 4.8 & 4.2 \\
47 & LibEST & RQ17\_FGUC\_006.txt & 4 & 4.1 & 5 & 4 & 4.2 & 5 \\
48 & LibEST & RQ19\_FGUC\_001.txt & 4 & 4.2 & 5 & 4 & 4.1 & 5 \\
49 & LibEST & RQ20\_FGUC\_001.txt & 4.3 & 4.7 & 5 & 4.4 & 4.6 & 5 \\
50 & LibEST & RQ23\_FGUC\_001.txt & 4.5 & 4.4 & 5 & 4.7 & 4.4 & 5 \\
51 & LibEST & RQ24\_FGUC\_001.txt & 4.8 & 4.3 & 4 & 4.5 & 4.6 & 4.3 \\
52 & LibEST & RQ26\_FGUC\_003.txt & 4.7 & 4.5 & 5 & 4.6 & 4.7 & 5 \\
53 & LibEST & RQ27\_FGUC\_001.txt & 4.4 & 4.2 & 5 & 4.3 & 4.5 & 5 \\
54 & LibEST & RQ36\_FGUC\_001.txt & 4.6 & 4.7 & 5 & 4.4 & 4.3 & 5 \\
55 & LibEST & RQ39\_FGUC\_001.txt & 4.5 & 4.5 & 5 & 4.5 & 4.3 & 5 \\
56 & LibEST & RQ46\_FGUC\_002.txt & 4.7 & 4.5 & 4.1 & 5.6 & 4.7 & 4.2 \\
57 & LibEST & RQ45\_FGUC\_001.txt & 4.7 & 4.4 & 5 & 4.5 & 4.4 & 5 \\
58 & LibEST & RQ49\_FGUC\_001.txt & 4.4 & 4.5 & 5 & 4.3 & 4.5 & 5 \\
59 & LibEST & RQ55\_FGUC\_001.txt & 4.6 & 4.2 & 5 & 4.4 & 4.6 & 5 \\
60 & LibEST & RQ57\_FGUC\_002.txt & 4.5 & 4.6 & 5 & 4.7 & 4.5 & 5 \\
\bottomrule
\end{tabular}
}
\end{table*}

\subsection*{Appendix B LLM‑based Automatic Generation of CNL‑B‑Formatted Normalized Use‑Case Specifications: Code Augmentation Based on Ground‑Truth Links}
\label{app:code_enhance}
\addcontentsline{toc}{subsection}{Appendix B LLM‑based Automatic Generation of CNL‑B‑Formatted Normalized Use‑Case Specifications: Code Augmentation Based on Ground‑Truth Links}

\begin{figure*}[t]
\centering
\caption{Code‑Augmentation Approach}
\label{fig9}
\includegraphics{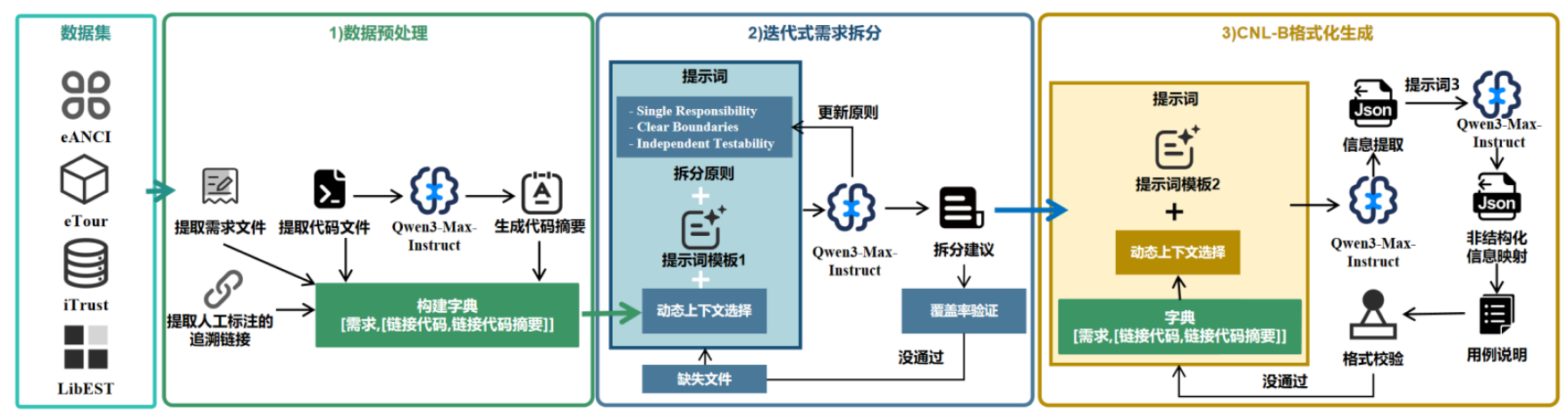}
\end{figure*}

To compare the performance difference between use‑case specifications generated from requirements only and those generated by feeding requirements together with code context based on ground‑truth traceability links, and to explore the effectiveness of code information for use‑case specifications, this study designs a code‑augmentation approach (see Figure \ref{fig9}). The framework includes three core stages: 1) Data preprocessing, which enhances the model’s semantic perception through code summarization techniques; 2) Iterative requirement splitting, which decomposes raw requirements into independent sub‑requirements via an iterative optimization mechanism; 3) CNL‑B formatted generation, which converts sub‑requirements into standardized CNL‑B format. It should be noted that this framework is solely intended to investigate the behavior of use‑case specifications under complete code information and is not applicable in practical industrial deployment.

The use‑case generation task in this framework also adopts the Qwen3‑Max‑Instruct model and leverages the CoT prompting strategy to improve the model’s reasoning stability and output standardization.

\subsubsection*{B.1 Data Preprocessing}
\addcontentsline{toc}{subsubsection}{B.1 Data Preprocessing}

The preprocessing procedure (Algorithm \ref{alg:data_preprocessing}) mainly consists of three steps. First, extract raw requirements $R$, source code $C$, and ground‑truth traceability links $G_\mathrm{links}$ from each project as the baseline. Second, employ an LLM to generate functional summaries for source code. File summaries $\mathit{Absfile}_{,j}$, function signatures $\mathit{Sig}_j$, and function summaries $\mathit{FuncAbs}_j$ are assembled to enrich semantic representations. Finally, construct the dictionary $\mathcal{S}_\mathrm{data}$ (containing requirements $R$, traceability‑linked code $C$, and code summaries $\mathit{LinkedCode}$) based on available traceability links, which serves as the base input for subsequent processing stages.

\begin{algorithm}[!htbp]
\centering
\begin{algorithmic}[1]
\Require Raw dataset $\mathcal{D}$, ground‑truth traceability links $G$, large language model $\mathit{LLM}$
\Ensure Structured dataset $\mathcal{S}_{\text{data}}$

\State $\mathcal{S}_{\text{data}} \gets \emptyset$ \Comment{Initialize empty structure}

\For{\textbf{each} dataset $d \in \mathcal{D}$}
    \State $R, C \gets \text{LoadRequirementsAndCode}(d)$ \Comment{Load requirement and code files}
    \State $G_{\text{links}} \gets \text{ParseTraceabilityLinks}(d)$ \Comment{Parse traceability links}
    
    \State \Comment{Step 1: Process each code file}
    \State $C' \gets \emptyset$ \Comment{Store processed code information}
    \For{\textbf{each} code file $c_j \in C$}
        \State $\mathit{Sig}_j \gets \text{ExtractSignatures}(c_j)$ \Comment{Extract function signatures}
        \State $\mathit{Abs}_{\text{file},j} \gets \text{GenerateFileAbstract}(\mathit{LLM}, c_j)$ \Comment{Generate file summary}
        \State $\mathit{FuncAbs}_j \gets \text{GenerateFunctionAbstracts}(\mathit{LLM}, c_j, \mathit{Sig}_j)$ \Comment{Generate function summaries}
        
        \State $C'[c_j] \gets \bigl\{\text{code}: \text{content}(c_j), \text{code\_abstract}: \mathit{Abs}_{\text{file},j}, \text{functions}: \{\mathit{Sig}_j, \mathit{FuncAbs}_j\}\bigr\}$
    \EndFor
    
    \State \Comment{Step 2: Process each requirement file}
    \For{\textbf{each} requirement file $r_i \in R$}
        \State $\mathit{ReqContent} \gets \text{content}(r_i)$ \Comment{Obtain requirement content}
        \State $\mathit{LinkedCode} \gets \emptyset$
        
        \If{$r_i \in \text{domain}(G_{\text{links}})$}
            \For{\textbf{each} code file $c_j$ linked to $r_i$ in $G_{\text{links}}$}
                \If{$c_j \in C'$}
                    \State $\mathit{LinkedCode}[c_j] \gets C'[c_j]$
                \EndIf
            \EndFor
        \EndIf
        
        \State $\mathcal{S}_{\text{data}}[r_i] \gets \bigl\{\text{req\_file\_content}: \mathit{ReqContent}, \text{traceability\_Links}: \mathit{LinkedCode}\bigr\}$
    \EndFor
\EndFor

\Return $\mathcal{S}_{\text{data}}$
\end{algorithmic}
\caption{Data Preprocessing and Structuring}
\label{alg:data_preprocessing}
\end{algorithm}

\subsubsection*{B.2 Iterative Requirement Splitting}
\addcontentsline{toc}{subsubsection}{B.2 Iterative Requirement Splitting}

\begin{algorithm}[!htbp]
\centering
\begin{algorithmic}[1]
\Require Structured dataset $\mathcal{S}_{\text{data}}$ (from Algorithm \ref{alg:data_preprocessing}), large language model client $\mathit{LLM}$, maximum iteration count $K=3$
\Ensure Analysis results $\mathcal{A}_{\text{results}}$ containing use‑case splitting suggestions

\State $\mathcal{P} \gets \text{InitialSplittingPrinciples}$ \Comment{Initialize splitting principles}
\State $\mathcal{A}_{\text{results}} \gets \emptyset$

\For{\textbf{each} entry $(r_i, C_{\text{links}}) \in \mathcal{S}_{\text{data}}$}
    \State $req\_content \gets r_i$
    \State $codes, abstracts \gets C_{\text{links}}$
    \If{$\text{EvaluateContextLimit}(req\_content, codes)$} \Comment{Evaluate whether context exceeds limit}
        \State $code\_ctx \gets codes$
    \Else
        \State $code\_ctx \gets abstracts$
    \EndIf
    
    \State $iter \gets 0$, $\text{is\_complete} \gets \mathbf{False}$, $\text{feedback} \gets \emptyset$
    \While{$iter < K$ \textbf{and not} $\text{is\_complete}$}
        \State $iter \gets iter + 1$
        \State $Prompt \gets \text{AssemblePrompt}(req\_content, code\_ctx, \mathcal{P}, \text{feedback})$ \Comment{Assemble prompt}
        \State $Response \gets \mathit{LLM}.\text{Generate}(Prompt)$
        \State $Result_i, \mathcal{P}_{\text{new}} \gets \text{Parse}(Response)$ \Comment{Parse response}
        \State $\mathit{Missing} \gets \text{ValidateTraceability}(Result_i, codes)$ \Comment{Check coverage omissions}
        \If{$\mathit{Missing} = \emptyset$}
            \State $\text{is\_complete} \gets \mathbf{True}$
            \State $\mathcal{P} \gets \text{Update}(\mathcal{P}, \mathcal{P}_{\text{new}})$ \Comment{Update splitting principles}
        \Else
            \State $\text{feedback} \gets \text{GenerateFeedback}(\mathit{Missing})$ \Comment{Generate feedback messages}
        \EndIf
    \EndWhile
    \State $\mathcal{A}_{\text{results}}[r_i] \gets Result_i$
\EndFor
\Return $\mathcal{A}_{\text{results}}$
\end{algorithmic}
\caption{Iterative Requirement Decomposition}
\label{alg:iterative_decomposition_2}
\end{algorithm}

The core task of this stage is to generate sub‑requirement splitting suggestions $\mathcal{A}_\mathrm{results}$ for raw requirements by leveraging the reasoning capability of the LLM under the guidance of decomposition criteria. To this end, this study designs and implements an iterative requirement decomposition algorithm, as illustrated in Algorithm \ref{alg:iterative_decomposition_2}.

When processing the structured dataset $\mathcal{S}_\mathrm{data}$ produced by Algorithm \ref{alg:data_preprocessing}, the system first adopts a dynamic code‑context selection strategy. According to the scale of the requirement to be split and its associated code, it determines whether full source code or code summaries shall be fed into the prompt, so as to balance context‑window constraints and information completeness.

Subsequently, the algorithm enters an iterative loop. Within each iteration, the system constructs a complete prompt by integrating splitting guidelines, the current requirement, and the selected code context, and submits it to the LLM. The generated splitting suggestions undergo coverage validation. By comparing splitting outputs against the original traceability matrix, the algorithm ensures that all associated code elements are covered.

If coverage omissions are detected, the missing code elements together with existing splitting outputs are composed into a new prompt and sent to the LLM to regenerate splitting suggestions. This procedure runs at most three times. Upon successful validation, an evolution mechanism is triggered: project‑specific characteristics summarized in the current round are absorbed to update decomposition criteria on‑the‑fly, improving the model’s adaptability to diverse project characteristics.

At initialization, the algorithm sets the following core decomposition principles:
\begin{itemize}
    \item Single responsibility: clear business operations actively initiated by actors
    \item Well‑defined boundaries: explicitly specified inputs, processing procedures, and outputs
    \item Independent and testable: testable without relying on other sub‑requirements
\end{itemize}

\subsubsection*{B.3 CNL‑B Formatted Generation}
\addcontentsline{toc}{subsubsection}{B.3 CNL‑B Formatted Generation}

To standardize the descriptions of use‑case specifications, this study adopts CNL‑B as the output format. Built upon the sub‑requirement splitting suggestions $\mathcal{A}_\mathrm{results}$ produced by Algorithm \ref{alg:iterative_decomposition_2}, this generation process is implemented via a three‑step pipeline, as shown in Algorithm \ref{alg:cnl_formatting}.

First, guided by splitting suggestions, the system assembles sub‑requirement descriptions and code context into prompts. The LLM is employed for information extraction to identify key entities including Actors, Core Operations, and Exceptions. During this procedure, full source code or code summaries are dynamically selected according to current context length to balance information completeness under model context‑window constraints. Structured use‑case generation follows, where extracted unstructured information is mapped onto the standard CNL‑B JSON template, establishing systematic mappings among requirement elements (actors, use‑cases, scenarios).

Finally, a format‑validation‑and‑retry mechanism automatically parses generated CNL‑B artifacts and inspects key fields (e.g., whether the main scenario is missing). Regeneration is triggered immediately upon validation failure to guarantee syntactic and structural compliance of output files.

\begin{algorithm}[!htbp]
\centering
\begin{algorithmic}[1]
\Require Analysis results $\mathcal{A}_{\text{results}}$ (from Algorithm \ref{alg:iterative_decomposition_2}), dataset structure $\mathcal{S}_{\text{data}}$, large language model client $\mathit{LLM}$
\Ensure Formatted CNL‑B use‑case files $\mathcal{F}_{\text{cnl}}$

\State $\mathcal{F}_{\text{cnl}} \gets \emptyset$
\State $P_{\text{extract}}, P_{\text{map}} \gets \text{LoadPromptTemplates}()$ \Comment{Load prompt templates}

\For{\textbf{each} project entry $(r_{file}, SubReqs) \in \mathcal{A}_{\text{results}}$}
    \State $C_{\text{links}}, Abs_{\text{links}} \gets \mathcal{S}_{\text{data}}[r_{file}].\text{traceability\_Links}$ \Comment{Retrieve traceability links}
    
    \For{\textbf{each} $(u_{id}, u_{desc}) \in SubReqs$}
        \State $C_{\text{target}} \gets \text{RetrieveCode}(u_{id}, C_{\text{links}})$ \Comment{Retrieve associated code}
        
        \If{\textbf{not} $\text{EvaluateContextLimit}(P_{\text{extract}}, u_{desc}, C_{\text{target}})$}
            \State $code\_ctx \gets C_{\text{target}}$ 
        \Else
            \State $code\_ctx \gets Abs_{\text{links}}$ 
        \EndIf
        
        \State $\mathit{Entities} \gets \mathit{LLM}.\text{Generate}(P_{\text{extract}}, u_{desc}, code\_ctx)$
        \State $\text{JSON}_{\text{cnl}} \gets \mathit{LLM}.\text{Generate}(P_{\text{map}}, \mathit{Entities})$
        
        \State $Result, isValid \gets \text{ParseAndVerify}(\text{JSON}_{\text{cnl}})$ \Comment{Parse and validate}
        
        \If{\textbf{not} $isValid$}
            \State $\text{ReGenerate}(u_{id})$ \Comment{Trigger regeneration}
        \Else
            \State $\text{SaveToFile}(r_{file}, u_{id}, Result)$ \Comment{Save result to file}
            \State $\mathcal{F}_{\text{cnl}} \gets \mathcal{F}_{\text{cnl}} \cup \{Result\}$
        \EndIf
    \EndFor
\EndFor
\Return $\mathcal{F}_{\text{cnl}}$
\end{algorithmic}
\caption{CNL‑B Format Conversion Algorithm}
\label{alg:cnl_formatting}
\end{algorithm}

\end{document}